\documentclass[final,5p,times,twocolumn]{elsarticle}

\usepackage{amssymb}
\usepackage{nidanfloat}
\usepackage{mathtools}
\usepackage{amsmath,amssymb,amsfonts}
\usepackage{graphicx}
\usepackage{textcomp}
\usepackage{hyperref}
\usepackage{caption,amsmath,graphicx,url,amssymb,latexsym,epsfig,tabularx,setspace,multirow,threeparttable,longtable,pdflscape,tabu,comment,subfigure,textcomp,xparse}
\usepackage{todonotes}
\usepackage[noend]{algpseudocode}
\usepackage{algorithm}
\usepackage{xcolor,soul}
\usepackage{forest, tikz}
\usepackage{xspace}
 \usepackage{hhline} 
\usepackage{enumitem}
\usepackage{listings}
\usepackage{makecell}

\newcolumntype{R}{>{\raggedleft\arraybackslash}X}
\newcolumntype{Y}{>{\centering\arraybackslash}X}

\usetikzlibrary{decorations.pathreplacing}

\usepackage{sidecap}
\sidecaptionvpos{figure}{c}

\def\knn{\textit{KNN}\xspace}
\def\gpu{\textsc{HybridKNN-Join}\xspace}
\def\hgpu{\textsc{Hybrid-GPU}\xspace}
\def\hcpu{\textsc{Hybrid-CPU}\xspace}
\def\cpuonly{\textsc{CPU-Only}\xspace}
\def\cpuonlyrr{\textsc{CPU-Only-RR}\xspace}
\def\gpuonly{\textsc{GPU-Only}\xspace}

\def\bufferkdtree{\textsc{BufferKDTree}\xspace}
\def\kdtree{\textsc{KDTree}\xspace}

\def\classuniform{\textsc{Unif-}\xspace}
\def\uniforma{\textsc{Unif2D}\xspace}
\def\uniformb{\textsc{Unif4D}\xspace}
\def\uniformc{\textsc{Unif6D}\xspace}

\def\classexpo{\textsc{Expo-}\xspace}
\def\expoa{\textsc{Expo2D}\xspace}
\def\expob{\textsc{Expo4D}\xspace}
\def\expoc{\textsc{Expo6D}\xspace}

\def\gaia{\textsc{Gaia}\xspace}
\def\osm{\textsc{Osm}\xspace}

\journal{JPDC}

\begin{document}

\begin{frontmatter}



\title{Hybrid KNN-Join: Parallel Nearest Neighbor Searches \\Exploiting CPU and GPU Architectural Features}


\author{Michael Gowanlock\corref{cor1}}

\cortext[cor1]{Corresponding author. Michael Gowanlock, School of Informatics, Computing, and Cyber Systems (Building \#90) 1295 S. Knoles Dr., Flagstaff, AZ, 86011,  U.S.A. E-mail: michael.gowanlock@nau.edu}

\address{School of Informatics, Computing, and Cyber Systems, Northern Arizona University, Flagstaff, AZ, U.S.A., 86011}

\begin{abstract}
K Nearest Neighbor (\knn) joins are used in scientific domains for data analysis, and are building blocks of several well-known algorithms.  \knn-joins find the \knn  of all points in a dataset. This paper focuses on a hybrid CPU/GPU approach for low-dimensional \knn-joins, where the GPU may not yield substantial performance gains over parallel CPU algorithms. We utilize a work queue that prioritizes computing data points in high density regions on the GPU, and low density regions on the CPU, thereby taking advantage of each architecture's relative strengths. Our approach, \gpu, effectively augments a state-of-the-art multi-core CPU algorithm. We propose optimizations that $(i)$ maximize GPU query throughput by assigning the GPU large batches of work; $(ii)$ increase workload granularity to optimize GPU utilization; and, $(iii)$ limit load imbalance between CPU and GPU architectures. We compare \gpu to one GPU and two parallel CPU reference implementations. Compared to the reference implementations, we find that the hybrid algorithm performs best on larger workloads (dataset size and K).  The methods employed in this paper show promise for the general division of work in other hybrid  algorithms.

\end{abstract}




\end{frontmatter}

\section{Introduction}\label{sec:intro}

This paper studies the \knn self-join problem, which is outlined as follows: given a database, $D$, of points, find all of the $K$ nearest neighbors of each point. We focus on the self-join because it is a common task in scientific data processing workflows (e.g., within an astronomy catalog, find the closest five objects of all objects within a feature space~\cite{2013AJ....146...22Z}). \knn searches are used in many applications, such as the k-means~\cite{hartigan1979algorithm}, and Chameleon~\cite{karypis1999chameleon} clustering algorithms. Consequently, \knn searches have been well studied~\cite{Roussopoulos:1995,arya1998optimal,muja2014}, including algorithms targeting the Graphics Processing Unit (GPU)~\cite{nam2016}.

There are several \knn research thrusts in the literature. \knn searches are employed in both low and high dimensional contexts. An example \knn query in the low dimensional case is as follows: find the closest $K$ restaurants to my current position, where the feature vectors contain 2-dimensional coordinates of the locations of nearby restaurants.  An example \knn search in the high dimensional context is image classification~\cite{kim12012comparing}, where image pixel intensities are converted to feature vectors, which may contain hundreds or thousands of features.      

To find the \knn of each point (or feature vector) in a dataset, one option is to perform a brute force search between all data points, which yields a quadratic complexity. Another option is to use an indexing data structure (e.g., kd-tree~\cite{Bentley:1975:MBS:361002.361007}, or R-tree~\cite{Guttman-R_tree}), which prunes the search for points which are nearby a given query point (we refer to a query point as a point being searched to find its $K$ nearest neighbors). In low dimensionality, the indexing data structures perform quite well and are able to discern between data points in each dimension, which reduces the quadratic complexity of the brute force algorithm~\cite{Roussopoulos:1995,arya1998optimal,muja2014,nam2016}.  

The  high dimensional context leads to several problems concerning the ``curse of dimensionality''~\cite{bellman1961}. In higher dimensionality, index searches typically become more exhaustive, where a \knn search for a given query point needs to compare to a substantial fraction of the points in the entire dataset. Thus, index searches become ineffective, and may even degrade performance relative to a brute force search, because searching the index incurs some degree of overhead. The exhaustive nature of high dimensional \knn searches led to the development of approximate algorithms~\cite{arya1998optimal,muja2014} that return $K$ nearby neighbors of a given query point, but they may not be the exact $K$ nearest neighbors. 

In this work, we focus on exact \knn searches in low dimensionality. The performance of low dimensional \knn searches is limited by the memory bottleneck.  However, the high aggregate memory bandwidth of modern GPUs~\cite{Volta} results in roughly an order-of-magnitude increase in memory bandwidth over the CPU. Therefore, GPUs are well-suited to data-intensive workloads. However,  data transfers to and from the GPU are a well-known bottleneck, which can decrease the performance advantages that are potentially afforded by the GPU. Additionally, many data-dependent workloads, such as the \knn-join studied in this work, have irregular execution patterns that make the GPU unsuitable for the algorithm due to thread divergence and serialization that degrades performance~\cite{Han:2011:RBD:1964179.1964184}. Thus, it is not clear that the GPU will lead to performance gains over multi-core CPU algorithms designed for \knn searches in low dimensionality.

Despite the great potential of GPU-accelerated \knn algorithms, much of the literature focuses on optimizing brute force approaches which highlight performance in high dimensional feature spaces and often compute a distance matrix~\cite{Garcia2010,arefin2012gpu,jian2013parallel,komarov2014fast}. The key idea is to compute the distance between a query point and all other points in $D$, then select the $K$ neighbors with the smallest distances to the query point.

We depart from the distance matrix approach, and focus on low dimensional \knn searches within a dimensionality regime that can employ indexing data structures to prune the search for potential neighbors of each query point in the dataset. Given the above context, we summarize the goals of the paper as follows.

\noindent\textbf{Addressing Low-Dimensionality on the GPU: } The abovementioned brute force \knn searches in high-dimensional feature spaces are clearly well-suited to the GPU as the many independent distance calculations can easily exploit the GPU's massive parallelism.  But, it is not clear that the GPU can significantly outperform parallel CPU approaches in low dimensionality. We address \knn searches in up to 6-D, which is largely the  domain of CPU \knn algorithms that employ indexing data structures. 
  
\noindent\textbf{Transforming the GPU-Accelerated Similarity Join into the KNN-Join: }  Recent work has proposed a similarity self-join for the GPU that finds all points within a search distance $\epsilon$ of a query point using an index~\cite{gowanlock2019accelerating}. The similarity join can be used to construct part of a \knn search by searching within a distance $\epsilon$ of a query point, and if there are $\geq K$ neighbors within $\epsilon$, order the neighbors by distance and select the nearest $K$ neighbors.  We leverage an efficient GPU similarity join algorithm in our approach.


\noindent\textbf{Concurrent Exploitation of CPU and GPU Resources: } In contrast to GPU-only approaches, we use both the CPU and GPU by assigning query points to either architecture to find their respective \knn. We leverage the distance similarity join described above for the GPU to process high data density regions, and a parallel CPU \knn algorithm for processing low density regions.

To our knowledge, our algorithm is the first  to split \knn searches between architectures as described above. The contributions and primary findings of the paper are outlined as follows:

\begin{itemize}
\item We propose a hybrid CPU/GPU approach for solving the \knn self-join problem that combines a distance similarity join for the GPU with a multi-core CPU \knn algorithm.
\item The GPU component of our \gpu algorithm solves the \knn problem using distance similarity searches. We show how to select a search distance, $\epsilon$, such that the GPU join is likely to find at least $K$ neighbors for each query point.
\item We present a work queue to distribute query points to the CPU and GPU. The work queue prioritizes assigning query points with significant computation to the GPU.
\item The throughput-oriented GPU requires processing large quantities of query points in batches to achieve peak performance. This can lead to load imbalance between the CPU and GPU. We propose methods to mitigate load imbalance between architectures. 
\item We compare the hybrid approach to one GPU and two multi-core CPU reference implementations. We find that \gpu outperforms the reference implementations on most scenarios, particularly at larger values of $K$.
\end{itemize}

The paper is organized as follows. Section~\ref{sec:background} presents background material. Section~\ref{sec:self_join_gpu} recaps leveraged GPU self-join literature. Section~\ref{sec:knn_join} presents the hybrid \knn self-join and optimizations. Section~\ref{sec:experimental_eval} evaluates our approach. And finally, Section~\ref{sec:conclusions} concludes the paper and discusses future work directions.

\section{Background}\label{sec:background}
\subsection{Problem Statement}
The \knn self-join is outlined as follows. Let $D$ be a database of $n$-dimensional points (or feature vectors) denoted as $p_i\in D$, where $i=1, 2, \dots,|D|$. For each point in the database, $p_i\in D$, we find its $K$ nearest neighbors, excluding the point itself. To compute the distance between two points, $p_a$ and $p_b$, we use the Euclidean distance as follows: $dist(p_a, p_b)=\sqrt{\sum_{j=1}^n(x^a_j-x^b_j)^2}$, where $x^a_j$ and $x^b_j$ denote the coordinates in dimension $j$ of point $a$ and $b$, respectively.    The \knn self-join is denoted as $D\bowtie_{KNN}D$. However, the \knn self-join problem and optimizations are also directly applicable to the case where there are two datasets $R$ and $S$ that are joined, $R\ltimes_{KNN}S$.  Table~\ref{tab:notation} summarizes notation that is used throughout multiple sections of the paper (notation that is self-contained within a section is not reported in the table).

Many data analytic and machine learning algorithms assume that processing can occur entirely in-memory. We make the same assumption here, and do not consider out-of-core (disk-based) solutions. For \knn searches, the result set size increases with $K$; therefore, the result set typically has the largest memory footprint of the algorithm. Since the result set may exceed GPU global memory capacity, we batch the execution across multiple GPU kernels to ensure that global memory is not exceeded. This allows our algorithm to process result set sizes that would otherwise exceed global memory capacity. We assume that the entire dataset can fit within the global memory of the GPU because the size of the dataset is much smaller than the result set.

\begin{table}[!t]
\centering
\begin{footnotesize}
\caption{Algorithm components and summary of notation.}\label{tab:notation}
\begin{tabularx}{\columnwidth}{|l|X|} \hline
\multicolumn{2}{|c|}{Names of Algorithms/Components}\\ \hline
\gpu     & The proposed CPU/GPU approach.\\\hline
\hcpu    & Parallel CPU component of \gpu.\\\hline
\hgpu    & GPU component of \gpu.\\\hline
\multicolumn{2}{|c|}{Notation}\\ \hline
$D$            &The input dataset.\\\hline
$p_i\in D$     &Data points in the dataset.\\\hline
$n$            &The dimensionality of the data.\\\hline
$K$            &The number of nearest neighbors found for each $p_i \in D$.\\\hline
$\epsilon$     &The search distance for \hgpu that may be dynamically expanded.\\\hline
$\epsilon_{min}$     &The initial search distance selected by \hgpu.\\\hline
$n_{large}$    &Initial monolithic batch size for \hgpu.\\\hline
$n_{small}$    &Small batch size for \hgpu.\\\hline
$n_{Cwin}$     &Window size of reserved queries for \hcpu during monolithic batch rounds.\\\hline
$n_{CPU}$      &Batch size for \hcpu.\\\hline
$t$            &Number of threads assigned to each query point for \hgpu.\\\hline
\end{tabularx}
\end{footnotesize}
\end{table}

\subsection{Related Work}
We present an overview of several categories of related work below.

\subsubsection{Hybrid Algorithms} A significant fraction of research on GPU algorithms and applications compares CPU vs. GPU approaches.  However, many GPU algorithms are unsuitable for all application scenarios, where parallel CPU algorithms may outperform the GPU in some instances. Consequently, using both multi-core CPUs and the GPU is needed to achieve peak performance in heterogeneous systems (see~\cite{Mittal2015} for a survey of hybrid algorithms). Thus, hybrid CPU/GPU algorithms  aim to maximize resource utilization in computer systems, and/or are designed such that they can take advantage of the relative strengths of each architecture. This paper is one such example of a hybrid algorithm that both splits the work between the CPU and GPU which maximizes resource utilization and also assigns work to each architecture to exploit the CPU and GPU's relative strengths.

Several works split the work between the CPU and GPU at runtime.  For instance, Li~et~al.~\cite{Li2011} parallelize Cryo-EM 3D reconstruction, and assign tasks to the CPU or GPU depending on the workload.  Deshpande~et~al.~\cite{Deshpande2011} filter images based on the degree of parallelism that varies across image regions, where the GPU is assigned the highly parallel regions and the CPU is assigned the remaining regions. Similarly to the abovementioned works, \gpu dynamically schedules the query points onto the architecture most suitable for the workload. To our knowledge, our preliminary work~\cite{Gowanlock2019GPGPU} on hybrid \knn-joins is the only such \knn algorithm that uses a hybrid approach. This comprehensive paper extends our preliminary work~\cite{Gowanlock2019GPGPU}.

\subsubsection{\knn Searches and Joins}
\knn searches are a fundamental machine learning algorithm. Consequently, there have been many works on optimizing the \knn search and join~\cite{Roussopoulos:1995,arya1998optimal,xia2004gorder,andoni2006near,yu2007efficient,yao2010k,Garcia2010,arefin2012gpu,leite2012nearest,jian2013parallel,gieseke2014buffer,komarov2014fast,muja2014,nam2016,patwary2016panda,ryoo2018efficient}. In this section, we discuss the related work on \knn searches. For clarity, we note that we only consider those algorithms that are capable of performing \emph{exact} searches. Approximate \knn searches are typically employed at higher dimensionality than that addressed in this paper.

We describe several \knn algorithms designed for the CPU as follows. An R-tree is used to find the \knn in~\cite{Roussopoulos:1995} that prunes the search for nearby candidate points to a given query point. The algorithm performs a branch-and-bound tree traversal, which first gets an estimate of the \knn and then performs backtracking on subtrees to find the exact neighbors. Backtracking in tree-based solutions~\cite{Roussopoulos:1995} is used to ensure that at least $K$ nearest neighbors are found.

The Approximate Nearest Neighbors (ANN) algorithm can be used to efficiently find both the approximate and the exact neighbors~\cite{arya1998optimal}. Approximate solutions are motivated by prohibitively expensive high-dimensional exact \knn searches. Related to ANN is the Fast Library for ANN (FLANN)~\cite{muja2014}, which achieves good performance using a parallel search over a randomized kd-forest. While FLANN outperforms ANN for one scenario in~\cite{muja2014}, the comparison was between a parallel (FLANN) and sequential (ANN) algorithm. Since ANN is considered a state-of-the-art exact \knn algorithm, we parallelize and incorporate it into \gpu.

There have been several efforts on parallelizing \knn searches on the GPU. We omit discussing the distance matrix based approaches~\cite{Garcia2010,arefin2012gpu,jian2013parallel,komarov2014fast} described in Section~\ref{sec:intro}, as we focus on lower dimensionality where indexing data structures are effective at pruning \knn searches.
A \knn GPU implementation is presented by Nam et al.~\cite{nam2016} that employs an R-tree index. Their algorithm performs backtracking in the R-tree to guarantee that $K$ neighbors are found for each query point. The algorithm is optimized to reduce warp divergence that occurs when executing  branch instructions that are necessitated by tree traversals.

A grid-based indexing solution for 3-dimensional \knn searches on the GPU was proposed by Leite et al.~\cite{leite2012nearest}. For a given query point, the algorithm expands the number of grid cells searched to ensure that at least $K$ neighbors are found. While the approach in~\cite{leite2012nearest} has some similarity to the work in this paper (both use a grid-based index), the algorithm uses a query-centric approach that expands the search radius when $<K$ neighbors are found. In contrast, in this paper, we avoid using a query-centric approach and instead elect to execute all \knn searches in a given batch of queries with a fixed search distance to minimize warp divergence. Our batched execution relaxes the constraint that $\geq K$ neighbors need to be found by each query point computed by the GPU.

\emph{Reference Implementation: } A GPU \knn search algorithm that uses a buffer kd-tree has been proposed by Gieseke~et~al.~\cite{gieseke2014buffer}. Similarly to the R-tree algorithm described above~\cite{nam2016}, the buffer kd-tree algorithm~\cite{gieseke2014buffer} avoids several drawbacks of the GPU's architecture. In particular, they search for queries in batches that are co-located within the same leaf. The algorithm delays execution by waiting for sufficient work to be accumulated into a buffer before accessing leaf nodes. This improves the SIMT parallelism of the algorithm.  Their algorithm also focuses on improving the fraction of coalesced memory accesses by having threads within a warp access either consecutive or nearby memory addresses. In our experimental evaluation, we compare our hybrid algorithm to this GPU algorithm that we denote as \bufferkdtree.

Table~\ref{tab:related_work} categorizes related work on \knn searches by the target dimensionality of the algorithm, whether the algorithm uses an indexing data structure to prune the search, or uses a brute force approach, the accuracy of the search (exact or approximate), and the target architecture (CPU, GPU or hybrid CPU/GPU). From Table~\ref{tab:related_work}, we observe that the majority of the low/moderate dimensionality works use indexing schemes, whereas the high dimensional algorithms focus on brute force approaches. As described in Section~\ref{sec:intro}, this is because the efficacy of indexing schemes to prune the search degrades at high dimensionality.

\begin{table*}[!t]
\centering
\begin{footnotesize}
\caption{Categorization of several \knn search algorithms in the literature. Categories: dimensionality, use of indexing methods, brute force searches, exact and approximate solutions, and architecture.}\label{tab:related_work}
\begin{tabularx}{\textwidth}{|l|X|X|l|l|l|l|l|l|X|} \hline
Reference & Low to Moderate Dimensionality & High Dimensionality & Indexing & Brute Force & Exact& Approx.&CPU & GPU & Hybrid CPU/GPU\\\hline
Roussopoulos et al.~\cite{Roussopoulos:1995}&X&&X&&X&&X&&\\\hline
Arya et al.~\cite{arya1998optimal}          &X&&X&&X&X&X&&\\\hline
Gieseke et al.~\cite{gieseke2014buffer}     &X&&X&&X&&&X&\\\hline
Nam et al.~\cite{nam2016}                   &X&&X&&X&&&X&\\\hline
Patwary et al.~\cite{patwary2016panda}      &X&&X&&X&&X&&\\\hline
Ryoo et al.~\cite{ryoo2018efficient}        &X&&X&&X&&&X&\\\hline
Leite et al.~\cite{leite2012nearest}        &X&&X&&X&X&&X&\\\hline
Gowanlock~\cite{Gowanlock2019GPGPU}         &X&&X&&X&&&&X\\\hline
Muja \& Lowe~\cite{muja2014}                &&X&X&&&X&X&&\\\hline
Garcia et al.~\cite{Garcia2010}             &&X&&X&X&&&X&\\\hline
Arefin et al.~\cite{arefin2012gpu}          &&X&&X&X&&&X&\\\hline
Jian et al.~\cite{jian2013parallel}         &&X&&X&X&&&X&\\\hline
Komarov et al.~\cite{komarov2014fast}       &&X&&X&X&&&X&\\\hline
\end{tabularx}
\end{footnotesize}
\end{table*}


\subsubsection{Indexing Techniques}
Central to our approach is using an appropriate index for the architecture. Indexes for the CPU have been designed to be work-efficient and data-aware, where spatial partitions are computed based on the input data. Examples include well-known tree-based indexes, such as kd-trees~\cite{Bentley:1975:MBS:361002.361007}, quad-trees~\cite{finkel1974}, and R-trees~\cite{Guttman-R_tree,prasad2015gpu}. In contrast, there are data-oblivious methods, such as statically partitioned grids~\cite{gowanlock2019accelerating}.

With the proliferation of general purpose computing on graphics processing units (GPGPU) there has been debate whether the community should use the tree-based approaches, or data-oblivious methods for the GPU. The disadvantage of index-trees is that searching the index using tree traversals requires performing many branch instructions, which can reduce the parallel efficiency of the GPU due to the SIMT architecture. Consequently, the abovementioned GPU \knn  algorithms that employed index-trees~\cite{gieseke2014buffer,nam2016} were optimized to avoid divergence. 

While there is little consensus regarding the type of index that should be employed on the GPU, we highlight two results. A GPU R-tree~\cite{Kim2015} was proposed and optimized to reduce thread divergence. Later, the same research group showed that it is better to perform the tree traversal on the CPU and perform the scanning of the leaf nodes on the GPU~\cite{KIM2018195}. This shows that the GPU should be leveraged through the use of regularized instructions, yielding low thread divergence. Therefore, we elect to use a non-hierarchical indexing technique for the GPU component of our \knn-join algorithm.

\subsubsection{Range Queries and Joins}
Our hybrid approach uses range queries (distance similarity searches) on the GPU to perform \knn searches. A join operation with a distance predicate can be implemented as several range queries. The multi-core CPU join algorithm in~\cite{kalashnikov2013} uses a non-materialized grid, and exploits the data distribution to efficiently perform a similarity join over a search distance, $\epsilon$, and the algorithm was shown to outperform the E$^2$LSH~\cite{andoni2006near}, and LSS~\cite{lieberman2008fast} algorithms.  A GPU self-join was presented in~\cite{gowanlock2019accelerating} that was shown to be efficient on low-dimensional data. We leverage some of the optimizations in the GPU self-join work~\cite{gowanlock2019accelerating} as they are effective for executing range queries that can be used to solve \knn searches on the GPU.

\subsubsection{Distributed Memory \knn Searches}
Distributed-memory approaches have been proposed to improve the performance of \knn searches. For instance, MapReduce~\cite{dean2008mapreduce} implementations for \knn joins~\cite{Lu:2012:EPK:2336664.2336674,zhang2012efficient} have been proposed. The authors in~\cite{Lu:2012:EPK:2336664.2336674} optimize the mapping function to prune distance calculations, which reduces the cost of  the shuffling operation and computation.  The authors in~\cite{zhang2012efficient} propose exact and approximate \knn join solutions, where they show that in their approximate solution, only a linear number of reducers are needed, which is a prerequisite for achieving good scalability. In contrast to scaling out the computation using distributed memory, we scale up the computation using the GPU.

\subsection{Application Scenario of this Paper}

There is a wide range of related work and application scenarios for \knn searches. In this paper, we focus on low to moderate dimensionality \knn searches (2--6 dimensions), where the \emph{curse of dimensionality}~\cite{bellman1961} does not prohibit indexing schemes from being effective at pruning \knn searches. This scenario is common in the literature (e.g., see the \knn search papers in Table~\ref{tab:related_work} by Roussopoulos et al.~\cite{Roussopoulos:1995}, Arya et al.~\cite{arya1998optimal}, Gieseke et al.~\cite{gieseke2014buffer}, Nam et al.~\cite{nam2016}, Patwary et al.~\cite{patwary2016panda}, Ryoo et al.~\cite{ryoo2018efficient}, and Leite et al.~\cite{leite2012nearest}). Additionally, we focus on \emph{exact} and not approximate searches, as the approximate searches are designed for much higher dimensionality than that considered in this paper. Furthermore, we reiterate that we depart from the literature by  splitting the work between CPU and GPU architectures.


\section{Recap of Previous Self-Join Work}\label{sec:self_join_gpu}


\gpu leverages the distance similarity self-join work of Gowanlock~\&~Karsin~\cite{gowanlock2019accelerating}, which was evaluated on up to $n=6$ dimensions. The authors used an efficient indexing scheme and batching scheme from~\cite{Gowanlock2017}, and proposed a technique to reduce the number of duplicate distance calculations. The approach was shown to outperform a state-of-the-art multi-core approach across many experimental scenarios; therefore, we employ their work in the GPU component of \gpu.  We outline the optimizations from~\cite{gowanlock2019accelerating}, that we use to efficiently solve the \knn-join on the GPU.

\subsection{Indexing Technique}\label{sec:indexing}
We use a grid-based indexing scheme for the GPU (see~\cite{gowanlock2019accelerating} for more detail) with cells of length $\epsilon$. The index is constructed on the host/CPU, and only stores non-empty grid cells, as indexing all cells may exceed the memory capacity of the GPU. The index, denoted as $G$, uses a series of lookup arrays to find relevant points in the index. A range query around a query point is carried out by performing distance calculations between points in each adjacent cell of the query point (and the cell containing the query point). The number of adjacent cells is $3^n$ (e.g., in 2-D there are 9 total adjacent grid cells). We make one minor change to the index described in previous work~\cite{gowanlock2019accelerating}, by removing the masking arrays, which were used to filter out cells that did not contain any points in a given dimension.  While the masking arrays may be useful in some scenarios (such as datasets with a bimodal distribution in a single dimension), in practice, we find that they had a negligible impact on performance. The space complexity of the index is $O(|D|)$. This small memory footprint allows for larger datasets and result set sizes to be processed on the GPU.        
While we use this grid-based indexing technique, we modify its construction, as discussed in Section~\ref{sec:indexing_on_GPU}.

\subsection{Batching Scheme}\label{sec:batching}
We give a brief overview of the GPU batching scheme in~\cite{gowanlock2019accelerating}. The size of the total result set for a join operation, which contains the neighbors of each point within a distance $\epsilon$, can be larger than the GPU's global memory capacity. To process large datasets or values of $\epsilon$, a batching scheme is needed to incrementally process the join, by querying a fraction of $D$ at each kernel invocation until range queries have been performed on all $p_i\in D$. We select a number of batches to execute by first estimating the total result set size (using a lightweight kernel), which yields an estimate, $e$, of the total result set size. Given a buffer size of $b_s$ (the size of a buffer to store the result set of a batch), we compute the total number of batches to be $n_b=\lceil e/b_s\rceil$. This obviates failure-restart strategies that can waste computation. We use 3 CUDA streams (a minimum of $n_b=3$), which overlaps the execution of the kernel and data transfers to exploit bidirectional PCIe bandwidth, and concurrent computation on the host. We use $b_s=10^8$ for each stream. 
In our experiments, on the larger workloads (higher dimensionality and $K$), using multiple streams is able to hide most of the host-GPU communication, whereas on the smaller workloads, the algorithm is bound by memory transfer operations.

\section{\gpu and Optimizations}\label{sec:knn_join}
\subsection{Splitting Work Between Architectures}
As discussed in Section~\ref{sec:intro}, we focus on a hybrid CPU/GPU approach that performs the \knn search using the CPU and GPU.   

A similarity search finds all points, $p_i\in D$, within a search distance, $\epsilon$, of a query point. Thus, to construct a \knn-join using a range query, there are several facets of the problem to consider. The $\epsilon$ search distance is required to ensure that the nearest points from a query point are found. For a given search that returns $>K$ neighbors, the distances between points are compared to determine which of the points are nearest to the query point. However, while a range query will return all points within $\epsilon$ of a batch of query points, there is no guarantee that all (or any) of the query points will have $K$ neighbors. In principle, the selection of $\epsilon$ could be large such that all points have at least $K$ nearest neighbors; however, this would lead to significant computational overhead, as some points in the dataset may find a large fraction of the entire dataset necessitating a significant number of distance calculations. 

\begin{figure}[!t]
\centering

\usetikzlibrary{decorations.pathreplacing}

\begin{tikzpicture}[scale=0.7]

\node at (0.5-3,3.75) {(a)};
\node at (5.5-2,3.75) {(b)};

\node at (2.0-2,-0.25) {\footnotesize{Dense Region: Good for the GPU}};
\node at (6.0,-0.25) {\footnotesize{Sparse Region: Good for the CPU}};


\foreach \x in {0,1,2} {
	\foreach \y in {0,1,2} {
		\draw [thick,black](\x-2,\y) rectangle (\x,\y+2);
	}
}


\foreach \x/\y in {2.4/2.2, 2.4/2.9, 3.2/2.2, 3.29/3.5, 2.25/2.7, 2.25/3.75, 2.36/0.15, 1.86/2.85, 1.5/3.9, 1.3/1.3, 1.3/1.8, 0.4/1.3, 3.4/1.2,3.4/1.2,3.1/1.6,2.5/1.8, 2.4/1.55} {
	\draw [fill=black] (\x-2,\y) circle (0.08cm);
}

\draw [fill=red] (2.2-2,2.5) circle (0.1cm);





\draw [thick,black](0+4,0) rectangle (4+4,2+2);


\foreach \x/\y in {3.4/3.2,3.1/3.6,2.5/3.8,2.4/3.55,3.6/3.7} {
	\draw [fill=black] (\x+4,\y) circle (0.08cm);
}

\draw [-, line width=1.0pt] (3.1+4,0) -- (3.1+4,4);

\draw [-, line width=1.0pt] (0+4,3.8) -- (3.1+4,3.8);
\draw [-, line width=1.0pt] (3.1+4,3.2) -- (4+4,3.2);

\draw [fill=red] (0.5+4,0.5) circle (0.1cm);


\end{tikzpicture}     
    \caption{Example query points assigned to either the GPU or CPU and possible indexing strategies for each. (a) The GPU is proficient at processing high density regions with a non-hierarchical grid. (b) The CPU is proficient for low density regions with an index-tree (kd-tree partitioning shown).}
   \label{fig:work_example}
\end{figure}

Figure~\ref{fig:work_example} shows an example of a spatially partitioned region with query points shown as larger red points. In Figure~\ref{fig:work_example}(a), there are many nearby neighbors to the query point; thus, a significant number of distance calculations need to be computed to find the $K$ nearest neighbors. However, in Figure~\ref{fig:work_example}(b), the query point is located in a sparse region. Thus, a large range query would be needed to find at least $K$ neighbors. Spatially partitioning the data using a grid in Figure~\ref{fig:work_example}(a) is reasonable, as it is likely $K$ neighbors will be found by checking adjacent cells (e.g., assume $K=3$). In contrast, in Figure~\ref{fig:work_example}(b), the grid is not effective. Had a grid been used, the adjacent cells would not contain any nearby points. In this case, a data-aware index (e.g., kd-tree~\cite{Bentley:1975:MBS:361002.361007} partitioning shown in Figure~\ref{fig:work_example}(b)) is better suited to finding data in sparse regions. Furthermore, as there are fewer points nearby the query point in Figure~\ref{fig:work_example}(b), there is a low degree of candidate point filtering overhead.       

Given this illustrative example, the GPU and associated indexing scheme in Section~\ref{sec:indexing} is good for processing the scenario in Figure~\ref{fig:work_example}(a) due to the large amount of filtering overhead needed (the massive parallelism of the GPU is well-suited to distance calculations), and low index search overhead; whereas the scenario in Figure~\ref{fig:work_example}(b) is good for finding the \knn on the CPU due to the low degree of filtering overhead and associated data-aware indexing scheme for low density regions. Therefore, the motivation for splitting the work between CPU and GPU is based on the suitability of each architecture to find the \knn of a given query point.

\subsection{Hybrid KNN-Join Overview}\label{sec:hybrid_overview}
We exploit the relative strengths of CPU and GPU architectures. The GPU is proficient at processing large batches of queries when the kernel can exploit the high memory bandwidth and massive parallelism afforded by the architecture. The CPU is better at processing irregular instruction flows, and thus, is well-suited to tree-based indexes that are comprised of many branch instructions.

\subsubsection{CPU KNN Component (\hcpu)}\label{sec:hcpu}
We use the publicly available\footnote{ANN can be found here: \url{http://www.cs.umd.edu/~mount/ANN/}.} ANN CPU implementation~\cite{arya1998optimal} that uses a kd-tree index. The algorithm is efficient for both approximate and exact solutions to the \knn problem, and we execute the algorithm such that we obtain the exact nearest neighbors. As noted in other work~\cite{patwary2016panda}, ANN uses global variables in its functions, which are not conducive to shared-memory parallelism. We obviate this limitation by parallelizing ANN using  MPI where the $K$ nearest neighbors of query points are found independently by each process rank. The results are written directly to an MPI shared memory window and thus we avoid explicit communication between process ranks.   We refer to the multi-core CPU approach of \gpu as \hcpu.

\subsubsection{GPU-Join Component (\hgpu)}\label{sec:hgpu}


In CPU-based \knn searches~\cite{Roussopoulos:1995}, backtracking is used to ensure that $K$ neighbors are found for each query point searched. Likewise, the E$^2$LSH~\cite{andoni2006near} CPU algorithm for range queries has been used for \knn searches by expanding the search radius until $\geq K$ neighbors are found for each point. As an example of expanding the search radius, Figure~\ref{fig:increase_eps_query_failure}(a) shows where $K=5$ neighbors are found when $\epsilon=1$, whereas Figure~\ref{fig:increase_eps_query_failure}(b) shows an example where $\epsilon$ needs to be expanded to $\epsilon=2$ to find at least $K=5$ neighbors. Backtracking or expanding the search radius is a query-centric approach that is beneficial for modern CPUs that can take advantage of the memory hierarchy (e.g., benefiting from locality during tree traversals), but is unsuitable for a batched GPU execution.    

\begin{figure}[!t]
\centering

\begin{tikzpicture}[scale=1.75]


\node at (-0.1,1.8) {(a)};

\foreach \x\y in {1.0/1,1.5/1.3,2.0/1.6} {
	\draw (1,1) circle (\x*0.5cm);
	\node at (1.12+0.5*\x,1) {\footnotesize{\x}};
}

\draw [fill=black!30!white](1,1) circle (0.5cm);

\filldraw [fill=red](1,1) circle (0.05cm);	

\foreach \x/\y in {0.3/1.1,1.2/1.3,0.8/0.8,0.9/1.8,1/0.2, 1.1/0.6, 1.33/0.84, 0.95/1.31} {
\filldraw [fill=black](\x,\y) circle (0.03cm);	
}


\node at (-0.1+2.5,1.8) {(b)};

\draw [fill=black!30!white](1+2.5,1) circle (1.0cm);

\foreach \x\y in {1.0/1,1.5/1.3,2.0/1.6} {
	\draw (1+2.5,1) circle (\x*0.5cm);
	\node at (1.12+0.5*\x+2.5,1) {\footnotesize{\x}};
}

\filldraw [fill=red](1+2.5,1) circle (0.05cm);	

\foreach \x/\y in {0.43/1.1,1.9/1.3,1.2/0.1,0.9/1.8,1/0.2, 1.1/0.6, 1.33/0.84, 0.95/1.31} {
\filldraw [fill=black](\x+2.5,\y) circle (0.03cm);	
}

\end{tikzpicture}     
    \caption{A \knn search around two query points (larger red points at the centers) where $K=5$. Shaded region denotes the range required to find $K=5$ points.  (a) $K=5$ neighbors are found with $\epsilon=1$. (b) $K=5$ neighbors are found when the search distance is expanded to $\epsilon=2$.}
   \label{fig:increase_eps_query_failure}
\end{figure}

To transform range queries with a distance $\epsilon$ into a \knn search that considers the throughput-oriented nature of the GPU, we use a batched execution that allows our GPU component, \hgpu, to fail to find at least $K$ points for each point searched. The overall idea that we will outline in Section~\ref{sec:workqueue}  is the following: $(i)$ the failed queries are added back to a work queue to be processed by either  \hgpu or \hcpu in the future; and $(ii)$ we dynamically re-index \hgpu with an increased $\epsilon$ value  when it reaches a threshold number of searches that did not yield $\geq K$ neighbors per point. Thus, each query point assigned to \hgpu is not guaranteed to find its \knn because we use a single $\epsilon$-distance when executing the kernel. Therefore, we refrain from using the query-centric approaches (e.g., backtracking, or increasing $\epsilon$ for individual point searches) on the GPU because this would lead to increased divergence in the kernel and intra-warp load imbalance.

\subsection{Algorithm Overview}
We present the pseudocode of \gpu to provide an overview of the technical details of the algorithm which are described later in this section. We outline \gpu in Algorithm~\ref{alg:knn_join} as follows.  Obtaining the process rank and importing the dataset occurs  on lines~\ref{algline.getrank}--\ref{algline.importdata}. We use an MPI implementation and have 1 master GPU rank and several CPU ranks which begin their primary execution on lines~\ref{algline.GPUstart}~and~\ref{algline.CPUstart}, respectively. For brevity, we do not show the work queue rank, as it simply assigns query points to the GPU and CPU ranks. 

The \hgpu rank initializes the result set (line~\ref{algline.resultkNNinit}), and query failure set (line~\ref{algline.failuresetinit}). Next, the value of $\epsilon_{min}$ is selected (Section~\ref{sec:select_epsilon}) on line~\ref{algline.selecteps}, and then $\epsilon$ is set using this value on line~\ref{algline.setepsilonfrommin} (we use $\epsilon_{min}$ later, which is why we declare both $\epsilon$ and $\epsilon_{min}$).  Next, we construct the index, $G$, as a function of $D$, and  $\epsilon$  on line~\ref{algline.constructindex}. Then, the algorithm gets a number of queries from the work queue rank on line~\ref{algline.getworkgpu1} and stores them in $Q_{GPU}$. A while loop is entered on line~\ref{algline.gpuwhile} that iterates until there are no more queries to compute (i.e., $|Q_{GPU}|=0$). Using the batch estimator, the number of GPU batches is computed on line~\ref{algline.numbatches} (recall from  Section~\ref{sec:batching} that the batch estimator computes the total number of batches so that \hgpu can process result sets larger than global memory). For clarity, note that these batches differ from the batches of queries obtained from the work queue ($Q_{GPU}$).  

The algorithm loops over all of the batches (line~\ref{algline.loopbatches}). At each iteration, the {\textsc{GPUJoinKernel}} is executed (line~\ref{algline.kernel}), which computes the result set for a single batch. On line~\ref{algline.result} the result of the join operation is filtered (the result is in the form of key/value pairs which are filtered to reduce duplicate keys), and store only points in $Q_{GPU}$ that have at least $K$ neighbors. On line~\ref{algline.filterknn}, those query points in the batch that have $<K$ neighbors are added to the $Q_{Fail}$ set, and these queries are added back to the work queue on line~\ref{algline.addfailtoqueue}.

On lines~\ref{algline.ifreindex}--\ref{algline.constructindex2}, the algorithm will dynamically re-index \hgpu with a larger $\epsilon$ value if $\geq25\%$ of points in $Q_{GPU}$ found $<K$ neighbors (Sections~\ref{sec:hgpu}~and~\ref{sec:workqueue}). And finally, on line~\ref{algline.getworkgpu2}, the rank retrieves work for the next batch from the work queue, and the $Q_{Fail}$ buffer is reset on line~\ref{algline.resetfailbuffer2}.

Regarding \hcpu, on line~\ref{algline.getworkcpu1}, queries are obtained from the work queue rank. If there are queries to process, a while loop is entered on line~\ref{algline.cpuwhile}, which computes the result of the \knn search for its batch of queries on line~\ref{algline.exactann}. The next batch of work is obtained from the work queue rank on line~\ref{algline.getworkcpu2}, and the loop continues until their are no additional queries to compute.

\begin{algorithm}[!t]
\caption{\gpu Algorithm}
\label{alg:knn_join}
\begin{algorithmic}[1]

\begin{footnotesize}
\Procedure{\gpu}{$K$, $b_s$}
\State myRank $\leftarrow$ getRank()\label{algline.getrank}
\State $D \leftarrow$ importData()\label{algline.importdata}
\If {myRank $=$ GPU Master Rank} \Comment{GPU Rank}\label{algline.GPUstart}
\State KNNresult $\leftarrow \emptyset$\label{algline.resultkNNinit}
\State $Q_{Fail}$ $\leftarrow \emptyset$\label{algline.failuresetinit}
\State $\epsilon_{min} \leftarrow$ selectEpsilon($D$)\label{algline.selecteps}
\State $\epsilon \leftarrow \epsilon_{min}$ \label{algline.setepsilonfrommin}
\State $G \leftarrow$ constructIndex($D$, $\epsilon$)\label{algline.constructindex}
\State $Q_{GPU}$ $\leftarrow$ getWork()\label{algline.getworkgpu1}
\While {$|Q_{GPU}|>0$}\label{algline.gpuwhile}
\State $n_b \leftarrow$ computeNumGPUBatches($b_s$, $Q_{GPU}$, $\epsilon$)\label{algline.numbatches}
\For {$i \in$  1, 2, $\dots$,$n_b$} \label{algline.loopbatches}
\State kernResult[i] $\leftarrow$ GPUJoinKernel($D$, $Q_{GPU}$, $G$, $\epsilon$, $i$)\label{algline.kernel}
\State KNNresult $\leftarrow$ KNNresult $\cup$ filterKeys(kernResult[i])\label{algline.result}
\State $Q_{Fail}$ $\leftarrow$ $Q_{Fail}$ $\cup$ findFailedPnts(kernResult[i], $Q_{GPU}$)\label{algline.filterknn}
\EndFor

\State addFailuresToWorkQueue($Q_{Fail}$)\label{algline.addfailtoqueue}
\If {$|Q_{Fail}|$/$|Q_{GPU}|>0.25$}\label{algline.ifreindex}
\State $\epsilon\leftarrow \epsilon+0.5\epsilon_{min}$\label{algline.increaseeps}
\State $G \leftarrow$ constructIndex($D$, $\epsilon$)\label{algline.constructindex2}
\EndIf
\State $Q_{GPU}$ $\leftarrow$ getWork()\label{algline.getworkgpu2}
\State $Q_{Fail}$ $\leftarrow$ $\emptyset$ \label{algline.resetfailbuffer2}
\EndWhile

\Else \Comment{CPU Ranks} \label{algline.CPUstart}
\State $Q_{CPU}$ $\leftarrow$ getWork()\label{algline.getworkcpu1}
\While {$|Q_{CPU}|>0$}\label{algline.cpuwhile}
\State KNNresult $\leftarrow$ KNNresult $\cup$ \hcpu($Q_{CPU}$, myRank) \label{algline.exactann}
\State $Q_{CPU}$ $\leftarrow$ getWork()\label{algline.getworkcpu2}
\EndWhile

\EndIf
\State \Return
\EndProcedure

\item[]

\Procedure{GPUJoinKernel}{$D$, $Q_{GPU}$, $G$, $\epsilon$, $i$}
\State resultSet $\leftarrow \emptyset$\label{algline.resultsetinit}
\State gid $\leftarrow$ getGlobalId($i$)\label{algline.globalid}
\State queryPoint $\leftarrow$ getPoint(gid, $Q_{GPU}$)\label{algline.getpoint}
\State adjCells $\leftarrow$ getAdjCells($G$, queryPoint)\label{algline.getadjcells}
\For {cell $\in$ adjCells.min,$\dots$,adjCells.max }\label{algline.loopcells}
\State pntResult $\leftarrow$ pntResult $\cup$ calcDistancePts(queryPoint, cell, $\epsilon$)\label{algline.calcdist}
\EndFor
\State resultSet $\leftarrow$ resultSet $\cup$ pntResult\label{algline.resultset}

\State \Return resultSet\label{algline.return}
\EndProcedure

\end{footnotesize}
\end{algorithmic}
\end{algorithm}

We describe the \hgpu join kernel, but refer the reader to~\cite{gowanlock2019accelerating} for more detail.  We make two minor changes to the self-join kernel to accommodate \gpu. First, we add a query set, as we do not want to compare all points to each other, as range queries are only needed for those points in $Q_{GPU}$. Second, we allow multiple threads to process an individual point (Section~\ref{sec:threads_workloads}). In the GPU join kernel shown in Algorithm~\ref{alg:knn_join}, the result set is initialized (line~\ref{algline.resultsetinit}), and then the global thread id is computed (line~\ref{algline.globalid}). Next, the query point assigned to the thread is stored (line~\ref{algline.getpoint}), and a loop iterates over all adjacent cells (lines~\ref{algline.getadjcells}--\ref{algline.loopcells}). The point assigned to the thread is compared to all points in the adjacent cells, where a result is stored when a point is found to be within $\epsilon$ of the query point (lines~\ref{algline.calcdist}--\ref{algline.resultset}). The result is stored as key/value pairs, where the key is the query point id, and the results are both the point id within $\epsilon$ of the key, and the distance between the points.      

If more than one thread computes the distance between a query point and points in neighboring cells, then each thread only computes a fraction of the points in the cell on line~\ref{algline.calcdist} (Section~\ref{sec:threads_workloads}).

\subsection{\hgpu:  Selecting the Search Distance }\label{sec:select_epsilon}
The input parameter to a \knn search is $K$; but \hgpu needs an $\epsilon$-distance which is expected to find at least $K$ neighbors for each point.  Analytically deriving $\epsilon$ is feasible when the input data distribution is known. However, real-world datasets have data distributions that make an analytical approach intractable.

Consider a search distance, $\epsilon_{min}$, that \emph{on average} finds $K$ neighbors per $p_i \in D$. Therefore, some points will find $\geq K$ neighbors, and some will find $<K$ neighbors. We derive $\epsilon_{min}$ which is used as an initial search distance for \hgpu.

We rely on the execution of two GPU kernels that sample the dataset to determine a good value of $\epsilon$. First, we simply sample 0.1\% of $D$, and compute the mean distance between points, denoted as $\epsilon_{mean}$.  Next, we define a number of bins, $n_{bins}$, that store the frequency of the distances between pairs of points that fall within the distance bin, where the width of each bin is $\epsilon_{mean}/n_{bins}$. We then select a sample of the points in the dataset and compute the distance between each of these points and every other point in $D$, and store the distances in the respective bin, where any distance $>\epsilon_{mean}$ is not stored (using a search distance of $\epsilon_{mean}$ will return a large fraction of the dataset; much larger than any reasonable value of $K$).\footnote{In the experimental evaluation, the datasets range from $10^7$ and $2.5\times10^7$ data points. We simply sample 500 points regardless of the dataset size. Since each of these points is compared to every other point in the dataset, a small sample size is sufficient to compute with high accuracy the cumulative number of points in each bin.} We compute the cumulative number of points in each bin. Let $\mathcal{B}_d$ denote the distance bins, where $d=1, 2, \dots, n_{bins}$. Each $\mathcal{B}_d$ stores: $(i)$ its distance range denoted as $[\mathcal{B}_d^{start},\mathcal{B}_d^{end})$, where $\mathcal{B}_d^{start}=(d-1)\cdot(\epsilon_{mean}/n_{bins})$,  and $\mathcal{B}_d^{end}=d\cdot(\epsilon_{mean}/n_{bins})$; $(ii)$ the number of points found within its distance range $[\mathcal{B}_d^{start},\mathcal{B}_d^{end})$, denoted as $\mathcal{B}_d^n$; $(iii)$, and the cumulative number of points in the bin (including bins with points at lower distances), denoted as $\mathcal{B}_d^{c}$, where $\mathcal{B}_d^{c}=\sum_{a=1}^{d}\mathcal{B}_a^{n}$. This yields a relationship between the search distance and the average number of neighbors that will be found. $\epsilon_{min}$ corresponds to the query distance that yields $K$ cumulative neighbors, where $\epsilon_{min}=(\mathcal{B}_d^{start}+\mathcal{B}_d^{end})/2$, where $\mathcal{B}_{d-1}^{c}< K\leq \mathcal{B}_d^{c}$.

We select $\epsilon=\epsilon_{min}$, which on average finds $K$ neighbors for each searched point. Figure~\ref{fig:selection_epsilon_lowd} shows a 2-D example of a search within the grid, where the grid cell length is equal to the search radius and thus the search is bound to adjacent cells (Section~\ref{sec:indexing}).

\begin{figure}[!t]
\centering

\usetikzlibrary{decorations.pathreplacing}

\begin{tikzpicture}[scale=1]



\foreach \x in {0,1} {
	\foreach \y in {0,1} {
		\draw [black](\x,\y) rectangle (\x+2,\y+2);
	}
}



\draw [fill=red] (1.5,1.5) circle (0.1cm);
\draw [fill=none] (1.5,1.5) circle (1.0cm);

\draw [->,line width=1.0pt, scale=1] (1.5,1.5) -- (2.5,1.5);
\node at (2.5,1.3) {{$\epsilon_{min}$}};




\end{tikzpicture}     
    \caption{A 2-D example of the search radius $\epsilon_{min}$, which probabilistically contains $K$ neighbors per $p_i \in D$.}
   \label{fig:selection_epsilon_lowd}
\end{figure}

\subsection{Assigning Work using a Work Queue}\label{sec:workqueue}
The GPU should execute range queries for points in dense regions, and the CPU should perform the \knn search in sparse regions (Figure~\ref{fig:work_example}). We begin by estimating the total amount of work required to execute each $p_i \in D$. We repurpose the grid index that is sent to the GPU (Sections~\ref{sec:indexing}~and~\ref{sec:select_epsilon}) to estimate the total work. For each $p_i\in D$, we check the total number of points that are found within the point's grid cell. This information requires simply performing a scan over the GPU index's non-empty grid cell array. For each point found within a given cell, the total number of points found within the cell are assigned to each point as an approximation of the amount of work that will need to be computed for that point. Then, we sort this array in non-increasing order by the number of points in each cell. Since the number of points in a cell will trace the data density in the immediate region around each point, this yields an estimate of the total amount of work for each point. Alternatively, we could count the number of neighbors in each point's cell and the adjacent cells to obtain a more accurate estimate of the amount of work computed by each query point; however, this would require substantial work, and thus we employ the simple procedure outlined above to estimate the work required of each point.

The GPU is efficient at performing distance calculations in high density regions, and the CPU is efficient at computing the lower density regions. Figure~\ref{fig:workqueue} shows a work queue illustration, where an array $C$ stores the number of points within the cell of each point $p_i \in D$. For example, $p_{64}$ and $p_{53}$ both have 32 points in their cell. In contrast, the last point in $C$, $p_{27}$, only has a single point in its cell (itself). The work queue assigns \hcpu query points starting at $C[|D|]$ in decreasing order, and assigns \hgpu query points starting at $C[1]$ in increasing order. Thus, the queries assigned to the CPU progressively require more work, and the queries assigned to the GPU progressively require less work. Depending on the data distribution, \hgpu may only compute the \knn of a small fraction of $D$, but perform similar levels of work as \hcpu.

\begin{figure}[!t]
\centering

\begin{tikzpicture}[scale=0.45]

\def\scale{0.8}

\foreach \x in {1,2,3,4,5,6,7,8,9,10,11,12,13,14,15,16,17,18,19,20} {
		\draw (\x*\scale,1*\scale) rectangle (\x*\scale+1*\scale,1*\scale+1*\scale);	
}

\foreach \x in {1,2,3,4,5,6,7,8,9,10,11,12,13,14,15,16,17,18,19,20} {
		\draw (\x*\scale,2.5*\scale) rectangle (\x*\scale+1*\scale,2.5*\scale+1*\scale);	
}

\foreach \x in {1,2} {
		\node at (\x*\scale+0.5*\scale,0.5*\scale) {\footnotesize{\x}};
		\node at (\x*\scale+0.5*\scale,1.5*\scale) {\footnotesize{32}};
}

\node at (3*\scale+0.5*\scale,0.5*\scale) {\tiny{$\dots$}};

\foreach \x\y in {4/30,5/31,6/32} {
		\node at (\x*\scale+0.5*\scale,0.5*\scale) {\footnotesize{\y}};
		\node at (\x*\scale+0.5*\scale,1.5*\scale) {\footnotesize{32}};
}

\foreach \x\y in {7/33,8/34,9/35} {
		\node at (\x*\scale+0.5*\scale,0.5*\scale) {\footnotesize{\y}};
		\node at (\x*\scale+0.5*\scale,1.5*\scale) {\footnotesize{20}};
}

\node at (10*\scale+0.5*\scale,0.5*\scale) {\tiny{$\dots$}};

\foreach \x\y in {11/50,12/51,13/52} {
		\node at (\x*\scale+0.5*\scale,0.5*\scale) {\footnotesize{\y}};
		\node at (\x*\scale+0.5*\scale,1.5*\scale) {\footnotesize{20}};
}

\node at (14*\scale+0.5*\scale,0.5*\scale) {\tiny{$\dots$}};

\foreach \x in {15,16,17,18,19,20} {
		\node at (\x*\scale+0.5*\scale,1.5*\scale) {\footnotesize{1}};
}

\foreach \x\y in {15/95,16/96,17/97,18/98,19/99} {
		\node at (\x*\scale+0.5*\scale,0.5*\scale) {\footnotesize{\y}};
}

\node at (21*\scale+0.5*\scale,0.5*\scale) {\footnotesize{$|D|=100$}};

\node at (0*\scale+0.35*\scale,1.5*\scale) {\footnotesize{$C$}};

\node at (0*\scale+0.35*\scale,3*\scale) {\footnotesize{$p_i$}};

\foreach \x\y in {1/64,2/39,4/3,5/60,6/53,7/82,8/94,9/61,11/77,12/14,13/93, 15/45,16/51,17/37,18/32,19/24, 20/27 } {
		\node at (\x*\scale+0.5*\scale,3.0*\scale) {\footnotesize{\y}};
		
}

\draw [ultra thick, ->] (1.0*\scale,-0.5*\scale) -- (5.0*\scale,-0.5*\scale);
\node at (3*\scale,-1.3*\scale) {\hgpu};

\draw [ultra thick, <-] (17.0*\scale,-0.5*\scale) -- (21.0*\scale,-0.5*\scale);
\node at (19*\scale,-1.3*\scale) {\hcpu};






\end{tikzpicture}     
    \caption{Example of a work queue with $|D|=100$ data points.  An array, $C$, stores the number of points within each cell for each $p_i \in D$. $C$ is sorted in non-increasing order, where \hgpu is assigned points with the greatest amount of work, and \hcpu is assigned points with the least amount of work.}
   \label{fig:workqueue}
\end{figure}

We outline several work queue performance considerations as follows.
\begin{itemize}
\item \emph{Load Imbalance --} Performance degrades while one architecture (CPU or GPU) waits for the other to finish processing their queries. 
\item \emph{Work Queue Overhead --} While the smallest work unit (a single query point) would lead to the best load balancing, there is overhead when accessing a work queue, and thus assigning batches of queries reduces work queue overhead. This is independent of the architecture requesting work to compute.
\item \emph{Maintaining GPU Throughput --} The GPU requires large batches of queries to maintain high query throughput, as executing a single query point on the GPU will underutilize its resources. In contrast, the CPU does not suffer from this limitation. 
\end{itemize}

The work queue performance considerations are similar to the classical trade-off between load imbalance and work queue overhead (e.g., static vs. dynamic scheduling of for loops in OpenMP~\cite{chandra2001parallel}). However, this scenario is different than this classical scenario, as the GPU requires larger query batches than the CPU to maintain high throughput. This can negatively impact load balancing, as the GPU may be assigned a large batch of points to compute towards the end of the computation, which would leave the CPU cores idle while waiting for the GPU to complete its work.

We propose several design decisions for the work queue to mitigate load imbalance while maintaining high GPU query throughput. 

We allow \hgpu to be assigned two types of batches: $(i)$ large monolithic batches containing a substantial fraction of $p_i \in D$; and, $(ii)$ small batches. For a derived $\epsilon$ value (Section~\ref{sec:select_epsilon}), \hgpu may not find the \knn for each point assigned to it (Section~\ref{sec:hgpu}). Each $p_i\in D$ that fails to find its \knn is added back to the work queue, and may be found by either \hgpu (when $\epsilon$ is expanded) or \hcpu in the future. At each monolithic batch round, we reduce the batch size by a factor of two.  We denote $n_{large}$ to be the size of the monolithic batch as a fraction of $|D|$.

A drawback of the monolithic batches is that \hgpu can request many query points to compute and starve the CPU (\hcpu) of work. Consequently, we implement a window of reserved query points for the CPU to compute during monolithic batch processing. Thus, each time the GPU requests a monolithic batch, the work queue manager determines the maximum number of GPU points that can be assigned to \hgpu, such that the CPU has at least a minimum number of points to compute. We denote the size of the fraction $|D|$ points reserved for \hcpu as $n_{Cwin}$. 

Using $n_{Cwin}$, and the fraction $|D|$ points that have already been processed by the CPU and GPU, denoted as $n_{Cproc}$ and $n_{Gproc}$, respectively,  if we let $n_{large}^l$ be the size of the monolithic batch at round $l$, then the size at round $l+1$ is as follows:   
\begin{equation*}
n_{large}^{l+1}{=}\mathrm{min}\big[0.5n_{large}^{l},\mathrm{max}\big(0,1-n_{Gproc}-n_{Cproc}-n_{Cwin}\big)\big].
\end{equation*}

Therefore, the monolithic batch size at round $l+1$ is either half the size of the monolithic batch at $l$, or a smaller size, as a function of the fraction of queries already computed and the window of reserved queries, until $n_{large}=0$. 

Once the monolithic batch size decreases to $n_{large}=0$, \hgpu reverts to smaller batches and no queries are reserved for \hcpu ($n_{Cwin}=0$), such that: $(i)$ the GPU is still utilized; and, $(ii)$ the GPU and CPU finish their computation at similar times. However, there may be a substantial number of queries to compute despite (potentially) executing several monolithic batches, as the CPU window will have reserved queries from being added to monolithic batches. We denote $n_{small}$ as the size of each smaller \hgpu (non-monolithic) batch, and $n_{CPU}$ as the size of each \hcpu batch, both given as a fraction of $|D|$.   

\hgpu may fail to find the \knn for many points if $\epsilon$ is not increased. As $C$ stores points from most to least work, with each processed GPU batch, there are more query points that fail to find their \knn. Thus, when using the small or monolithic batches, we dynamically re-index \hgpu by increasing $\epsilon$ by a distance of $\epsilon_{min}/2$, when on the previous batch, \hgpu failed to find the \knn of at least 25\% of its assigned points. This dynamic approach attempts to reach a trade-off between $(i)$ not increasing $\epsilon$ too much which is expensive; and, $(ii)$ not failing to find too many query points in the batch. Re-indexing occurs in parallel on the GPU to reduce the time where the GPU is idle due to expanding $\epsilon$. Finally, when 95\% of the query points have found their \knn, we then decrease the batch sizes assigned to the CPU and GPU to $n_{CPU}/2$ and $n_{small}/2$, respectively. These smaller batches (half of the initial size) mitigates load imbalance at the end of the computation.

Figure~\ref{fig:workqueue_monolithic} illustrates the monolithic batches from the work queue being assigned to \hgpu and small batches of queries assigned to \hcpu. Figure~\ref{fig:workqueue_monolithic}(a) shows an initial work queue, where $1/3$ of $D$ ($n_{large}=1/3$) is assigned to \hgpu, and $1/3$ of the queries must be reserved for the CPU ($n_{Cwin}=1/3$). In Figure~\ref{fig:workqueue_monolithic}(b), after \hgpu processes its queries from the first batch, some of the queries will be complete (the \knn were found for these query points) and some will have failed to find the \knn. Hence, because there are a mix of complete and incomplete queries, we show this as partially complete in the figure. The vertical lines denote $n_{large}$ (dashed line) and $n_{Cwin}$ (solid line). The CPU is guaranteed to find the \knn of each query point, thus the queries are shown as complete. Comparing Figure~\ref{fig:workqueue_monolithic}(a)~and~(b) we see that the maximum GPU batch size does not increase substantially because $n_{large}$ is halved between rounds. Comparing Figure~\ref{fig:workqueue_monolithic}(c)~and~(d), the window of reserved CPU queries decreases the queries available for \hgpu to compute using a monolithic batch.  After $n_{large}=0$, \hgpu reverts to smaller batches of size $n_{small}$. Table~\ref{tab:reconfiguration} summarizes algorithm reconfiguration operations that occur during execution.

\begin{table*}[!t]
\centering

\begin{footnotesize}
\caption{Summary of \gpu algorithm reconfiguration operations that occur at runtime.}\label{tab:reconfiguration}
\begin{tabularx}{\textwidth}{|X|X|X|X|} \hline
&When&Why&Where\\\hline
Increasing $\epsilon$ and Re-indexing&The number of failed queries on the previous batch exceeds 25\% of the queries assigned in that batch.&Re-indexing occurs because the search distance, $\epsilon$, is increased to reduce the number of failed searches.&GPU\\\hline
GPU Decreasing Batch Size: Monolithic Batches &At each batch assigned to the GPU.&The GPU is initially assigned a very large batch of work to compute ($n_{large}$). The batch size assigned to the GPU decreases at each batch to mitigate against load imbalance between the CPU and GPU until $n_{large}=0$ indicating that the monolithic batch rounds have completed.  &CPU (work queue) and GPU\\\hline
GPU Small Batch Sizes&At each batch assigned to the GPU after $n_{large}=0$.&The GPU executes smaller batches of size $n_{small}$. &CPU (work queue) and GPU\\\hline
CPU and GPU Smallest Batch Sizes&95\% of the queries have been completed.&CPU and GPU batch sizes are decreased to $n_{CPU}/2$ and $n_{small}/2$, respectively, to mitigate against load imbalance between the CPU and GPU.&CPU and GPU\\\hline
\end{tabularx}
\end{footnotesize}
\end{table*}

\begin{figure}[htp]
\centering

\usetikzlibrary{patterns}

\begin{tikzpicture}[scale=0.43]

\def\scale{0.8}


\draw[fill=white] (1*\scale,5*\scale) rectangle (1*\scale+1*\scale,5*\scale+1*\scale);	
\node at (5.2*\scale,5.5*\scale) {\footnotesize{Not Yet Assigned}};

\draw[pattern=crosshatch, pattern color=black] (9*\scale,5*\scale) rectangle (9*\scale+1*\scale,5*\scale+1*\scale);
\node at (13.3*\scale,5.5*\scale) {\footnotesize{Partially Complete}};

\draw[fill=black] (17.5*\scale,5*\scale) rectangle (17.5*\scale+1*\scale,5*\scale+1*\scale);
\node at (20.5*\scale,5.5*\scale) {\footnotesize{Complete}};

\node at (-0.7*\scale,3.5*\scale) {\footnotesize{Max.}};
\node at (-0.7*\scale,2.5*\scale) {\footnotesize{$n_{large}$}};
\node at (-0.8*\scale,1.5*\scale) {\footnotesize{(a) $1/3$}};
\node at (-0.8*\scale,-2.5*\scale) {\footnotesize{(b) $1/6$}};
\node at (-0.8*\scale,-6.5*\scale) {\footnotesize{(c) $1/12$}};
\node at (-0.8*\scale,-10.5*\scale) {\footnotesize{(d) $1/24$}};

\foreach \xa\xb\y in {1/22/1, 1/22/-3, 1/22/-7, 1/22/-11} {
		\draw (\xa*\scale,\y*\scale) rectangle (\xb*\scale,\y*\scale+1*\scale);	
}

\draw [ultra thick, ->] (1.0*\scale,0.5*\scale) -- (5.0*\scale,0.5*\scale);
\node at (3*\scale,-0.3*\scale) {\hgpu};

\draw [ultra thick, <-] (18.0*\scale,0.5*\scale) -- (22.0*\scale,0.5*\scale);
\node at (20*\scale,-0.3*\scale) {\hcpu};

\draw [dashed] (8*\scale,0*\scale) -- (8*\scale,3*\scale);

\draw[decorate,decoration={brace,amplitude=5pt}] 
    (1*\scale,2*\scale) node(t_k_unten){} -- 
    (8*\scale,2*\scale) node(t_k_opt_unten){};
\node at (4.165*\scale,2.4) {\footnotesize{Max. GPU Batch Size}};

\draw [solid] (15*\scale,0*\scale) -- (15*\scale,3*\scale);

\draw[decorate,decoration={brace,amplitude=5pt}] 
    (15*\scale,2*\scale) node(t_k_unten){} -- 
    (22*\scale,2*\scale) node(t_k_opt_unten){};
\node at (18.66*\scale,2.4) {\footnotesize{CPU Window}};


\foreach \xa\xb\y in {1/7/-3} {
		\draw[pattern=crosshatch, pattern color=black] (\xa*\scale,\y*\scale) rectangle (\xb*\scale+1*\scale,\y*\scale+1*\scale);	
}

\foreach \xa\xb\y in {17/21/-3} {
		\draw[fill=black] (\xa*\scale,\y*\scale) rectangle (\xb*\scale+1*\scale,\y*\scale+1*\scale);	
}

\draw [solid] (10*\scale,-1*\scale) -- (10*\scale,-4*\scale);

\draw [dashed] (9*\scale,-1*\scale) -- (9*\scale,-4*\scale);

\foreach \xa\xb\y in {1/8/-7} {
		\draw[pattern=crosshatch, pattern color=black] (\xa*\scale,\y*\scale) rectangle (\xb*\scale+1*\scale,\y*\scale+1*\scale);	
}

\foreach \xa\xb\y in {10/21/-7} {
		\draw[fill=black] (\xa*\scale,\y*\scale) rectangle (\xb*\scale+1*\scale,\y*\scale+1*\scale);	
}

\draw [solid] (3*\scale,-5*\scale) -- (3*\scale,-8*\scale);

\draw [dashed] (3*\scale,-5*\scale) -- (3*\scale,-8*\scale);

\foreach \xa\xb\y in {1/8/-11} {
		\draw[pattern=crosshatch, pattern color=black] (\xa*\scale,\y*\scale) rectangle (\xb*\scale+1*\scale,\y*\scale+1*\scale);	
}

\foreach \xa\xb\y in {7/21/-11} {
		\draw[fill=black] (\xa*\scale,\y*\scale) rectangle (\xb*\scale+1*\scale,\y*\scale+1*\scale);	
}

\draw [solid] (1*\scale,-9*\scale) -- (1*\scale,-12*\scale);

\draw [dashed] (1*\scale,-9*\scale) -- (1*\scale,-12*\scale);









\end{tikzpicture}     
    \caption{Assigning monolithic batches of queries from the work queue to \hgpu and queries to \hcpu (small \hgpu batch rounds not shown). (a) Initial work queue with $n_{large}=n_{Cwin}=1/3$. (b) After processing a monolithic batch, some queries have been computed by \hgpu and \hcpu, and the monolithic batch size deceases. (c) The CPU window reduces the monolithic batch size. 
    (d) After processing with \hgpu the monolithic batch rounds are finished as $n_{large}=0$.}
   \label{fig:workqueue_monolithic}
\end{figure}

Note that we have made several parameter selection decisions. We dynamically re-index \hgpu when  25\% of queries fail to find at least $K$ neighbors in the previous batch. If we were to use a value $>$25\%, then the number of failed queries would increase, whereas if we use $<$25\% then the algorithm would spend more time re-indexing.  Furthermore, we use half of the small GPU batch sizes ($n_{small}$), and the CPU batch size ($n_{CPU}$)  when 95\% of the queries have found their \knn in the dataset to obviate load imbalance at the end of the computation. While these parameters are arbitrarily selected, we believe that they are reasonable design decisions (e.g., similarly, OpenMP guided scheduling reduces the chunk size with increasing iteration~\cite{chandra2001parallel}). In the experimental evaluation, we quantify the load imbalance between the CPU and GPU architectures, which demonstrates that the selection of parameters does indeed yield low load imbalance. The low load imbalance partially justifies the selection of these parameter values.

\subsection{GPU: Indexing}\label{sec:indexing_on_GPU}
As discussed in Section~\ref{sec:indexing}, we use a grid-based index for the GPU. In the previous work on the similarity self-join that we leverage, the index was constructed on the host/CPU, because we only needed to construct the index once. However, since finding the \knn on the GPU may require dynamically expanding $\epsilon$ several times, constructing the index on the host may become a bottleneck and degrade performance. Additionally, if the algorithm constructs the index on the host, then this reduces the resources available for the CPU component of the algorithm.  Consequently, while we use the index described in previous work in Section~\ref{sec:indexing}, we construct the index on the GPU. We find that index construction on the GPU is much faster than constructing the index on the host. Our preliminary work on \knn-joins~\cite{Gowanlock2019GPGPU}, that this paper extends, constructed the index on the host, and we found that it reduced performance particularly on lower values of $K$ (smaller workloads), where the ratio of re-indexing overhead to computation is high. On the larger workloads, re-indexing overhead on the host is amortized across the entire computation and only has a minor impact on performance.

\subsection{GPU: Optimizing Task Granularity}\label{sec:threads_workloads}
To improve resource utilization for parallel algorithms executed on the CPU or GPU, a common strategy is to reduce the size of each task to a smaller unit of work, and then redistribute these smaller tasks to threads or processes~\cite{Merrill:2012:SGG:2145816.2145832}. We describe a similar optimization as follows.  

In the self-join work that we leverage~\cite{gowanlock2019accelerating}, a single thread is assigned to each point in the dataset, where the thread finds all points within $\epsilon$ of its assigned point. This approach was tenable because the total number of threads is large ($|D|$). Since \hgpu may only process a small fraction of $D$ in a batch,  then the GPU's resources may be underutilized if we use one thread per point. Also, the GPU hides high memory latency by performing fast context switching between resident threads. Thus, oversubscribing the GPU by using more threads than cores is needed to saturate resources.

\begin{figure}[!t]
\centering

\usetikzlibrary{decorations.pathreplacing}

\begin{tikzpicture}[scale=0.55]


\foreach \x in {0,2,4} {
	\foreach \y in {0,2,4} {
		\draw (\x,\y) rectangle (\x+2,\y+2);
	}
}

\draw [dashed, line width=0.07cm, blue](4,4) rectangle (6,6);

\draw [->,line width=1.3pt, scale=1] (6,5) -- (6.9,3);

\foreach \x/\y in {4.8/4.8, 4.4/4.9, 5.2/4.2, 5.29/5.5, 4.25/4.7, 4.25/5.75} {
	\draw [fill=black] (\x,\y) circle (0.08cm);
}

\foreach \x/\y in {1.25/1.25,  4.36/2.15, 2.86/4.85, 3.2/5.8, 0.5/0.8, 0.3/1.2} {
	\draw [fill=black] (\x,\y) circle (0.08cm);
}
 %

\draw [fill=red] (3.6,3.7) circle (0.1cm);

\draw [->,line width=2.0pt, scale=1] (4,0) -- (6,0);
\node at (5.0,-0.25) {{\Large$\epsilon$}};



\node at (10.0,4.4) {{\small Points}};

\foreach \x/\y in {7/0,8/1,9/2,10/3,11/4,12/5} {
		\draw (\x,2.5) rectangle (\x+1,3.5);
		\node at (\x+0.5,3.0) {{\small$p_{\y}$}};
}


\node at (10.0,1.0) {{\small Thread ids}};

\foreach \x/\y in {7/0,8/1,9/2,10/0,11/1,12/2} {
		\node at (\x+0.5,2.25) {{\small$t_{\y}$}};
}


\draw[decorate,decoration={brace,amplitude=5pt}] 
    (7.2,3.75) node(t_k_unten){} -- 
    (12.8,3.75) node(t_k_opt_unten){};

\draw[decorate,decoration={brace,amplitude=5pt,mirror}] 
    (7.2,1.75) node(t_k_unten){} -- 
    (12.8,1.75) node(t_k_opt_unten){};

\end{tikzpicture}     
    \caption{Using multiple threads to compute the distances between points in 2-D.}
   \label{fig:threads}
\end{figure}

We divide the work of the distance calculations for a single point between multiple threads to increase task granularity. Figure~\ref{fig:threads} shows an example of using multiple threads per query point. The query point (red) is shown in the middle cell. The distances between the query point and the six points are computed in an adjacent cell (dashed blue outline). This example shows three threads each computing the distances between two points.




We assign a static number of threads per query point for performing the distance calculations, where the number of threads are referred to as $t$ (e.g., $t=32$ denotes using 32 threads per point).  An advantage of this approach is that the number of threads per point can be selected to reduce intra-warp thread divergence. For example, if $t=32$ threads per point are used, then a full warp will compute the distance between a given point and the candidate points. There should be low divergence  because each thread in the warp executes similar execution pathways. Recently, Gallet~and~Gowanlock~\cite{GalletGowanlock2019} showed that on the similarity self-join, computing the distances between a query point and candidate points using more than one thread improves warp execution efficiency. However, as discussed above, the primary motivation for using $t>1$ threads is that the \knn-join computes batches of query points which may be much smaller than the number of points in the dataset. Consequently, we need to use more threads per query point to saturate GPU resources.

Two drawbacks of using $t>1$ include: $(i)$ using too many threads per point may increase overhead; and $(ii)$ query points in lower density regions may not need a large number of threads, and such threads will have minimal work to compute. There is a trade off between assigning too few or too many threads per point. We assume that the number of threads selected to compute the distance calculations for each point should evenly divide the size of a warp (32 threads). This eliminates the possibility of the threads assigned to a point spanning multiple warps and increasing divergence.

\section{Experimental Evaluation}\label{sec:experimental_eval}

\subsection{Datasets}
We focus on low-dimensional \knn-joins due to their utility in many applications. Additionally, related work has consistently shown that GPU-accelerated \knn searches outperform CPU approaches at high dimensionality~\cite{Garcia2010,kato2012multi}, due to the increased cost of distance calculations. Thus, the GPU may be unsuitable to low-dimensional \knn-joins, and we target this low-dimensionality scenario.

We employ two classes of synthetic datasets with different workload characteristics. The \classuniform class of datasets contains uniformly distributed data points. The \classexpo class of datasets contains exponentially distributed data points with $\lambda=40$. Datasets are generated in 2, 4, and 6 dimensions for both classes, and contain $|D|=10^7$ points. Since \gpu splits the low and high density regions between the CPU and GPU, respectively, the \classuniform datasets represent the case where there is very low variation in density across the data space, whereas the \classexpo datasets represent the case where there is a large data density gradient. Comparing the performance between these two classes of datasets allows us to examine how performance may vary as a function of data distribution and workload assignment between the CPU and GPU.

We also employ two 2-D real-world datasets: \gaia which contains $|D|=2.5\times10^7$ positions of astronomical objects from the \textit{Gaia} catalog~\cite{brown2018gaia}, and \osm which contains $|D|=2.5\times10^7$ positions from Open Street Map data~\cite{osm}. Since the Open Street Map data contains GPS positions, we removed duplicate point coordinates, otherwise the \knn of many points may consist of a single GPS trajectory with identical (or nearly identical) point coordinates.

\subsection{Experimental Methodology}
All \gpu CPU code is written in C/C++, compiled using the GNU compiler (v.~5.4.0) with the O3 flag. The GPU code is written in CUDA v.~9. We use OpenMPI v.~3.1.1 for parallelizing host-side tasks (discussed in Section~\ref{sec:hcpu}). The work queue performs minimal work; however, we parallelize it using two OpenMP threads for assigning queries to \hcpu and \hgpu, as we need to wait on \hgpu without blocking \hcpu from obtaining new work.

Our platform consists of an NVIDIA GP100 GPU with 16 GiB of global memory, and has 2$\times$ Intel E5-2620~v4 2.1 GHz CPUs, with 16 total physical cores. The \hgpu kernel uses 256 threads per block. In the experiments, we exclude the time needed to load the dataset or construct the \hcpu indexes or the initial \hgpu index (see Section~\ref{sec:implementations}). The response time of performing the \knn search on the CPU and GPU is measured after the indexes have been constructed by \hcpu.  All other components of the algorithm are included in the response time (e.g., finding the search distance for \hgpu, and ordering the workload for the work queue). All time measurements are averaged over 3 trials. 

Table~\ref{tab:params} outlines default parameter values in the experimental evaluation. Note that the initial monolithic batch size ($n_{large}$) and the window size of reserved CPU queries ($n_{Cwin}$)  are both 40\% of $|D|$. Increasing $n_{large}$ beyond 40\% is unlikely to greatly improve performance as the larger the value of $n_{large}$, the more queries that fail to find $\geq K$ neighbors. $n_{small}$ and $n_{CPU}$ are 0.5\% of $|D|$, which is selected to minimize load imbalance, while not assigning too few queries per batch, which can increase the overhead of accessing the work queue.

\begin{table}[!t]
\centering

\begin{footnotesize}
\caption{Default parameter values used in the experimental evaluation.}\label{tab:params}
\begin{tabularx}{\columnwidth}{|X|X|} \hline
Parameter& Value\\\hline
$n_{large}$&$0.4|D|$\\\hline
$n_{small}$&$0.005|D|$\\\hline
$n_{Cwin}$ &$0.4|D|$\\\hline
$n_{CPU}$  &$0.005|D|$\\\hline
$t$        &8\\\hline
\end{tabularx}
\end{footnotesize}
\end{table}

\subsection{Implementations}\label{sec:implementations}
The implementations we use to carry out our performance evaluation are described below. 

\begin{itemize}
\item \cpuonly --  We compare to a multi-core CPU ANN~\cite{arya1998optimal} implementation that obtains the exact neighbors, as described in Section~\ref{sec:hcpu}. We compare \gpu to \cpuonly to demonstrate the performance gains yielded by the GPU. \cpuonly uses the CPU component of the hybrid algorithm, \hcpu, as executed with 16 processes that perform \knn searches, and 1 process for the work queue. There is no communication between ANN process ranks, as they find the \knn independently and write results directly to shared memory. Recall that we needed to parallelize ANN using MPI. We have each process rank independently construct its own kd-tree which is queried in batches obtained from the work queue. Since ANN does not perform parallel index construction and we cannot share the index between processes, we exclude this index construction time.     

\item \gpu -- Our hybrid approach uses: \hcpu with 15 processes, and \hgpu and the work queue each use 1 process. \hcpu uses one less process than \cpuonly. Due to the exclusion of the index construction time for \cpuonly (described above), we exclude the \hcpu index construction time, and the initial index constructed by \hgpu. However, we include all index construction time in the measurements when $\epsilon$ must be dynamically expanded by \hgpu.

\item \gpuonly -- We compare to a GPU-only implementation that uses \hgpu (the GPU component of \gpu).  \hgpu and the work queue each use 1 process. We note that because this implementation may fail to find at least $K$ neighbors by design, depending on the data distribution, it may take significant time to find the \knn of all points, as those points in sparse regions were intended to be found by the \hcpu component of \gpu. Thus, we show this implementation for comparison purposes, but note that it is not designed for standalone \knn searches. We configure \gpuonly to use only monolithic batches, i.e., $n_{large}=1$ and $n_{Cwin}=0$ for all batches computed by the algorithm. We ensure that each query batch size is equal to the number of points that have not yet found their respective \knn.

\item \cpuonlyrr -- We evaluate the potential negative performance impact of the work queue. We compare \cpuonly (described above) to another CPU implementation that has the work queue removed. Instead of using the work queue, we simply assign points in a round robin fashion to each process rank. Each $p_i \in D$ is assigned to rank $r$ when $[(i-1) ~\mathrm{mod}~ (P)]+1=r$, where $P$ is the number of processes (MPI ranks), and $r=1, 2, \ldots, P$. The implementation with the work queue removed uses $P=16$ processes; therefore, the number of ranks computing the \knn is equivalent to the \cpuonly implementation. We elect to use a round robin distribution of points to process ranks to achieve good load balancing.

\item \bufferkdtree -- We compare to the GPU buffer kd-tree algorithm by Gieseke et al.~\cite{gieseke2014buffer} discussed in Section~\ref{sec:background}. In all experiments where we compare to \gpu we set the tree depth parameter to 12, as it achieves good performance across several datasets (this will be demonstrated in Section~\ref{sec:bufferkdtree_height}). The algorithm is designed for the same low/moderate dimensional searches that we examine in this paper. To maintain consistency with the methodology in Gieseke et al.~\cite{gieseke2014buffer}, when reporting the response time, we only include the time to compute the query/test phase which finds the nearest neighbors. \bufferkdtree allows for searching up to $K=99$ neighbors; in our evaluation, the maximum value of $K$ tested on this algorithm is $K=96$. The \bufferkdtree code is publicly available~\cite{bufferkdtreecode}.

\item \kdtree -- We compare to the multi-core CPU implementation in Gieseke et al.~\cite{gieseke2014buffer}. The algorithm is configured to use 16 threads (the number of physical cores on our platform), and similarly to \bufferkdtree, we only report the query response time. The code is publicly available~\cite{bufferkdtreecode}.

\end{itemize}

\subsection{Results}

\subsubsection{Scalability of the \cpuonly Implementation}\label{sec:cpuonlyscalability}
\begin{figure}[!t]
\centering
\includegraphics[width=0.4\textwidth, trim={0.4cm 0.4cm 0.4cm 0.4cm}]{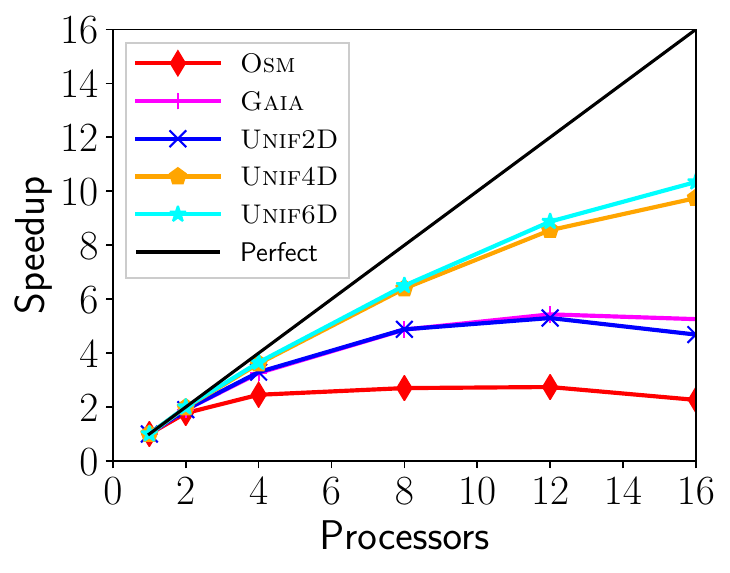}
    \caption{Speedup of the \cpuonly reference implementation on the \osm, \gaia, \uniforma, \uniformb, and \uniformc datasets where $K=32$.}
   \label{fig:cpuonly_scalability}
\end{figure}

The \cpuonly reference implementation is a parallelized version of ANN~\cite{arya1998optimal} and is used for the CPU component of the hybrid algorithm, \hcpu (Section~\ref{sec:hcpu}).  \cpuonly, uses 16 process ranks to independently find the \knn of batches of query points.   Figure~\ref{fig:cpuonly_scalability} plots the scalability of \cpuonly across several datasets where $K=32$. We find that scalability improves with data dimensionality. For example, \gaia is a 2-D dataset and achieves a speedup of 5.26$\times$ with 16 processes, whereas the 6-D \uniformc dataset achieves a speedup of 10.34$\times$ with 16 processes. The cost of the Euclidean distance calculation scales with dimensionality. Therefore, finding the \knn on lower dimensional datasets is memory-bound (the algorithm spends most of the time performing tree traversals), which transitions to becoming more compute-bound as the dimensionality increases. On \uniforma, the speedup slightly decreases from 12 to 16 processes. This is indicative of memory bandwidth saturation, where adding more processors does not improve performance (the same trend is observed on \gaia and \osm). However, on \uniformb and \uniformc, the greatest speedup is achieved when 16 processes are used, which indicates that memory bandwidth may not be saturated on those datasets when 16 processes are used. The poor scalability of the 2-D \osm dataset is surprising given that the other 2-D datasets (\gaia and \uniforma) achieve larger speedups.      

In summary, \cpuonly (and \hcpu) achieves good scalability on the higher dimensional datasets, but the speedup is limited on the low dimensional datasets.

\subsubsection{Scalability of the \kdtree Implementation}

\begin{figure}[!t]
\centering
\includegraphics[width=0.4\textwidth, trim={0.4cm 0.4cm 0.4cm 0.4cm}]{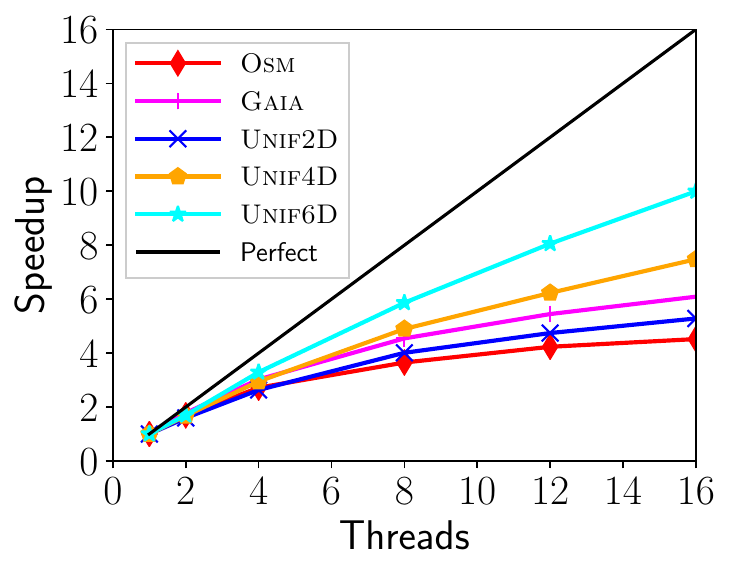}
    \caption{Speedup of the \kdtree reference implementation on the \osm, \gaia, \uniforma, \uniformb, and \uniformc datasets where $K=32$.}
   \label{fig:kdtree_scalability}
\end{figure}

The multi-threaded CPU \kdtree implementation provides another baseline for comparison. In contrast, \cpuonly is parallelized using MPI. We execute the same experiment in Section~\ref{sec:cpuonlyscalability} using \kdtree. Figure~\ref{fig:kdtree_scalability} plots the speedup vs. the number of threads. We find very similar scalability using \kdtree as we find for \cpuonly in Section~\ref{sec:cpuonlyscalability}. Since 16 threads achieves the best performance, we configure \kdtree to use 16 threads when we compare it to the other implementations.

\subsubsection{Impact of the \bufferkdtree Height Parameter}\label{sec:bufferkdtree_height}

\begin{figure}[!t]
\centering
\includegraphics[width=0.4\textwidth, trim={0.4cm 0.4cm 0.4cm 0.4cm}]{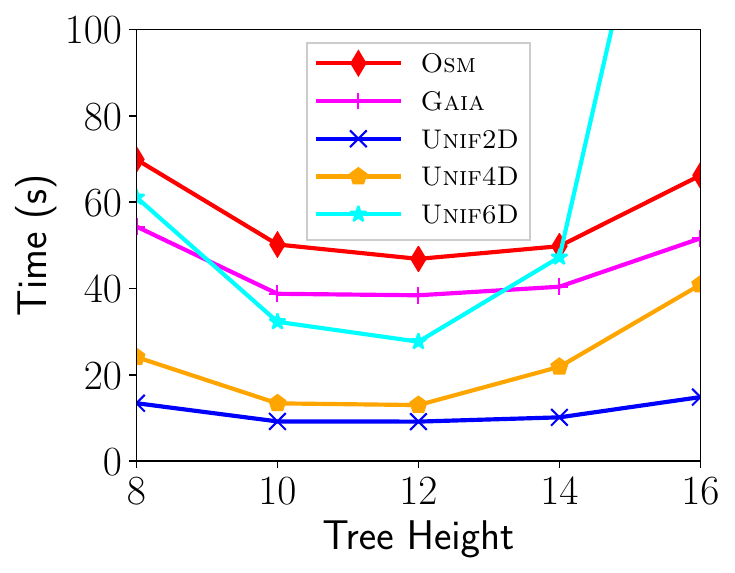}
    \caption{Impact of the \bufferkdtree tree height parameter on performance. The response time vs. tree height is plotted for the \osm, \gaia, \uniforma, \uniformb, and \uniformc datasets where $K=16$. The best performance is achieved across all datasets when the tree height is 12.}
   \label{fig:bufferkdtree_height}
\end{figure}

The \bufferkdtree implementation uses a height parameter that achieves a trade-off between examining too many leaves and pruning overhead caused by tree traversals. To ensure that we configure \bufferkdtree with a height parameter that achieves good performance, we execute \bufferkdtree on several datasets across different values of the height parameter. From Figure~\ref{fig:bufferkdtree_height}, we find that a height value of 12 achieves the best performance across all datasets, and this performance behavior is consistent with the experimental evaluation in Gieseke et al.~\cite{gieseke2014buffer}. In all future experiments, we use this tree height.     

\subsubsection{Potential Impact of Work Queue Overhead on Performance}\label{sec:workqueue_access_overlead_round_robin}
Utilizing work queues can add overhead to parallel algorithms (e.g., accesses must be serialized to avoid race conditions). We compare two parallel CPU-only algorithms: \cpuonly that uses the work queue, and \cpuonlyrr that has the work queue removed, where each point is assigned to a rank in a round robin fashion. By comparing the implementation with the work queue to the same implementation without the work queue, we are able to assess whether using the work queue adds considerable overhead.

\begin{figure}[!t]
\centering
\subfigure[\uniforma]{
\includegraphics[width=0.147\textwidth, trim={0.5cm 0.4cm 0.5cm 0.4cm}]{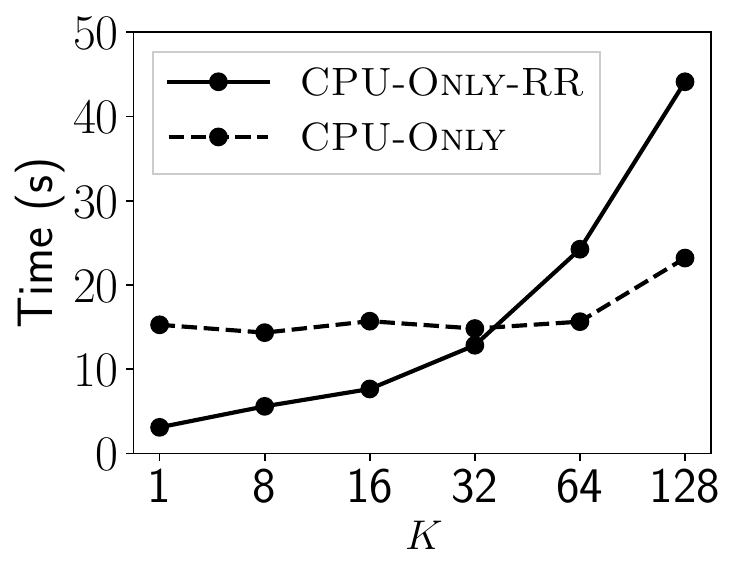}
}
\subfigure[\uniformb]{
\includegraphics[width=0.147\textwidth, trim={0.5cm 0.4cm 0.5cm 0.4cm}]{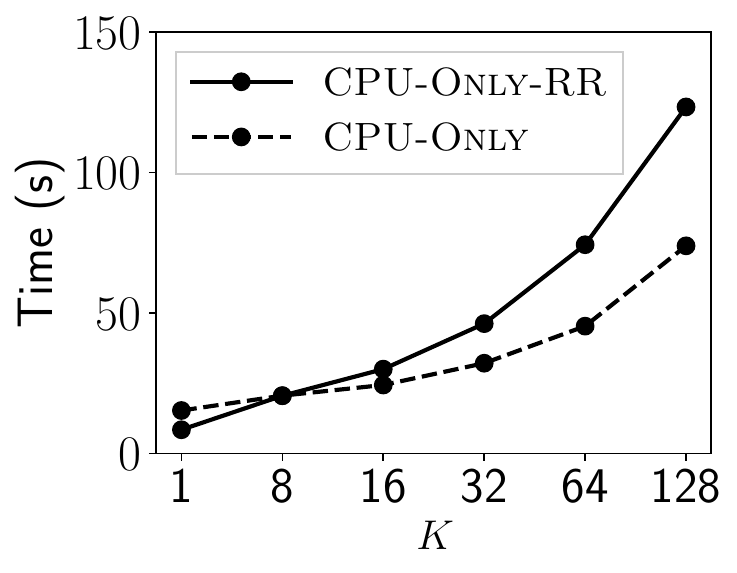}
}
\subfigure[\uniformc]{
\includegraphics[width=0.147\textwidth, trim={0.5cm 0.4cm 0.5cm 0.4cm}]{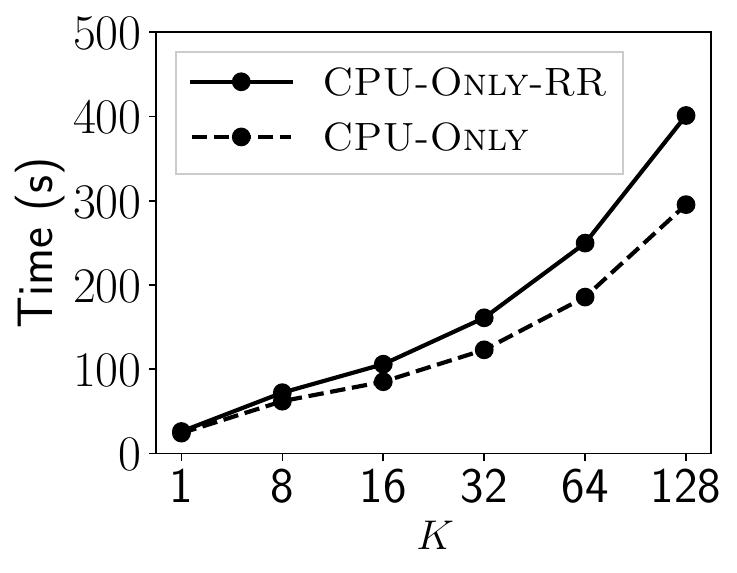}
}
    \caption{Response time vs. $K$ comparing \cpuonly and \cpuonlyrr on the uniformly distributed datasets.}
   \label{fig:time_vs_K_workqueue_on_off}
\end{figure}

Figure~\ref{fig:time_vs_K_workqueue_on_off} plots the response time vs. $K$ on the uniformly distributed datasets. We observe from the figure that the work queue improves performance over the round robin assignment of points on the larger workloads. For instance, when $K\geq8$ on the  \uniformb and \uniformc datasets, the implementation with the work queue (\cpuonly) outperforms the round robin distribution of points to ranks (\cpuonlyrr). However, on the \uniforma dataset, \cpuonlyrr  outperforms \cpuonly when $K\leq 32$. Because the workload is relatively low in 2-D, the initial work queue overheads have a non-negligible performance impact (we elaborate on these overheads in Section~\ref{sec:results_overheads}), and accessing the work queue has a non-negligible impact, where several processes contend for a new batch of work to compute.

Overall, we observe that the work queue generally has a positive impact on performance. In fact, the performance gain from the work queue over the round robin implementation is substantial in the cases described above (e.g., Figure~\ref{fig:time_vs_K_workqueue_on_off}(b)~and~(c)). This is an unintended benefit of the work queue. We attribute this to two factors, described as follows: 

\begin{enumerate}
\item The work queue helps reduce load imbalance between process ranks, as queries are retrieved from the queue on-demand.

\item The work queue first sorts the points by total work based on the number of points found in each cell. This means that spatially co-located points found in the same cell are likely to be assigned to the same process rank by the work queue. Because the points are spatially co-located, it is likely that the tree traversals are benefiting from good locality when performing \knn searches. In contrast, the round robin assignment of points to each process rank cannot benefit as much from the spatial co-location of points. 
\end{enumerate}

We test the notion that cache effects are improving the performance of \cpuonly over \cpuonlyrr described above, by simply using the Performance analysis tools for Linux (perf) to count the total number of cache references and misses using the \verb|cache-references| and \verb|cache-misses| flags. We note that perf yields a coarse grained measurement of the cache references and misses in the program, and is not limited to \knn searches. However, since the vast majority of the computation is performing \knn searches (tree traversals and filtering) in these algorithms, perf is a reasonable tool for understanding whether good locality is responsible for the performance difference between the \cpuonly and \cpuonlyrr implementations.

Table~\ref{tab:perf_cache_misses} shows the number of cache references and percentage of cache misses when executing the \cpuonly and \cpuonlyrr implementations on the \uniformb dataset. We find that the percentage of cache misses increases with $K$ in the \cpuonlyrr implementation, whereas this percentage decreases with $K$ in the \cpuonly implementation. We expect that the impact on locality of \cpuonly favors larger values of $K$. The larger the value of $K$, the larger the search space, which is likely to increase the probability of two searches for differing points to be able to reuse data in cache. In contrast, the \cpuonlyrr implementation processes queries in a round robin fashion and cannot exploit locality between consecutive searches. Therefore, as $K$ increases, the total percentage of cache misses increases.

\begin{table}[!t]
\centering
\begin{footnotesize}
\caption{Comparison of the total number of cache references and percentage of cache misses on the \cpuonly and \cpuonlyrr implementations on the \uniformb dataset. The lowest values of the percentage of cache misses as a function of $K$ are highlighted in bold face.}\label{tab:perf_cache_misses}
\begin{tabularx}{\columnwidth}{|l|R|R|R|R|} \hline
$K$&\cpuonly Cache References&\cpuonly Cache Misses (\%)&\cpuonlyrr Cache References&\cpuonlyrr Cache Misses (\%)\\ \hline
  1 &29208826489&\textbf{52.18}&26790397165 & 57.80\\\hline
  8 &32379870022&\textbf{54.42}&31912590674 & 64.11\\\hline
 16 &34398747794&\textbf{54.72}&35841307333 & 67.81\\\hline
 32 &37541677665&\textbf{53.76}&41829138108 & 72.19\\\hline
 64 &44754599329&\textbf{48.18}&51312595208 & 76.67\\\hline
128 &62648741911&\textbf{38.39}&68117485096 & 81.73\\\hline
\end{tabularx}
\end{footnotesize}
\end{table}

By comparing \cpuonly and \cpuonlyrr, we find that the work queue does not add significant overhead to the \cpuonly algorithm (with the exception of the \uniforma dataset when $K\leq 32$), and therefore, does not hinder \gpu. Furthermore, we find that assigning batches of query points to processes through the work queue has a positive effect on locality. The performance gain due to positive cache effects outweighs the performance loss due to work queue access overhead.

\subsubsection{Overheads of Work Queue Construction and Selection of the Search Distance}\label{sec:results_overheads}
Table~\ref{tab:workqueue_overhead} quantifies the fraction of the total response time of major overheads when $K=32$ on the \classuniform and \classexpo classes of datasets when executing \gpu with the default parameter values in Table~\ref{tab:params}. From Table~\ref{tab:workqueue_overhead}, we observe that the fraction of time computing $\epsilon$ (Section~\ref{sec:select_epsilon}) and ordering the work queue workload (Section~\ref{sec:workqueue}) is dependent on the data dimensionality rather than the data distribution, as the uniform and exponentially distributed datasets require similar fractions of the total response time for each of these overheads.

The percentage of the total response time to compute $\epsilon$  ranges from 6.3\% (\uniforma) to  1.2\% (\uniformc), and these percentages for ordering the work queue workload  range from 6.8\% (\uniforma) to 1.1\% (\uniformc). Thus, the overheads associated with constructing the work queue are mostly amortized on larger workloads; however, they are non-negligible on the smaller workloads. Despite these work queue construction overheads, from Section~\ref{sec:workqueue_access_overlead_round_robin} we find that the work queue is generally advantageous due to positive cache effects.

\begin{table}[!t]
\centering

\begin{footnotesize}
\caption{\gpu work queue construction overheads on the \classuniform and \classexpo classes of datasets where $K=32$. The total response time and the fraction of the total response time is shown for computing $\epsilon$ and ordering the work queue workload. The default parameter values in Table~\ref{tab:params} are used.}\label{tab:workqueue_overhead}
\begin{tabularx}{\columnwidth}{|l|R|R|R|} \hline
Dataset&Total response time (s)&Fraction computing $\epsilon$&Fraction ordering work queue workload\\ \hline
\uniforma& 9.82&0.063&0.068\\\hline
\uniformb&19.54&0.037&0.035\\\hline
\uniformc&74.50&0.012&0.011\\\hline
\expoa   &10.82&0.057&0.063\\\hline
\expob   &22.97&0.038&0.035\\\hline
\expoc   &77.21&0.013&0.011\\\hline
\end{tabularx}
\end{footnotesize}
\end{table}

\subsubsection{GPU Kernel Task Granularity}\label{sec:kernel_performance}
\hgpu uses a number of threads ($t$) to process each point (Section~\ref{sec:threads_workloads}). Since the size of each batch assigned from the work queue to \hgpu, $|Q_{GPU}|$, can vary (e.g., due to decreasing monolithic batch sizes), it is important that sufficient threads are executed, such that GPU resources remain saturated, which is achieved through oversubscription. Additionally, using more than one thread per query point allows for less divergence in each warp, as fewer query points (with differing execution pathways) are assigned to a single warp.

Table~\ref{tab:kernel_optimizations} shows the total response time of \gpu  for a selection of datasets, and values of $K$ (8, 32, 128) and $t$ threads (1, 8, 16, and 32) assigned to perform the distance calculations for each query point. We select values of $t$ such that the threads divide evenly into the size of a warp (32 threads). This ensures that a query point does not span two warps.  From Table~\ref{tab:kernel_optimizations} we observe that many of the datasets have consistent response times when varying $t$. For instance, on \gaia with $K=8$, the response times range from 19.90~s to 21.00~s. In contrast, on the \uniformc dataset with $K=32$, the response times range from 73.99~s to 90.01~s. Using $t=1$ may lead to inter-warp load imbalance (waiting for the last warp to finish execution) and increases divergence in the kernel. At the other extreme, using a single warp ($t=32$) to compute each query point may underutilize resources. For instance, if there are fewer than 32 candidates within an adjacent cell of a query point, then several threads will not have any work to execute.    Across all datasets and values of $t$, we find that $t=8$ yields the best performance. Additionally, in many cases where $t=8$ does not yield the best performance, we find that it achieves similar performance to the best performing value of $t$. Consequently, we configure \gpu with a default value of $t=8$.

\begin{table}[!t]
\centering
\begin{footnotesize}
\caption{Response time (s) of \gpu when varying $t$ for $t=1, 8, 16, 32$ and $K=8, 32, 128$ on \gaia, \osm, \uniformb, and \uniformc datasets. Excepting $t$, the default parameter values in Table~\ref{tab:params} are used. The lowest response time is shown in bold face for each $K$ on each dataset.}\label{tab:kernel_optimizations}
\begin{tabularx}{\columnwidth}{|c|R|R|R|R|R|} \hline
Dataset &$K$&$t=1$&$t=8$&$t=16$&$t=32$\\\hline
\gaia        &  8&     19.96&\bf{19.90}&     19.98&     21.00\\\hline
\osm         &  8&     36.31&     32.45&\bf{31.97}&     32.34\\\hline
\uniformb    &  8&\bf{11.75}&     12.23&     12.50&     14.00\\\hline
\uniformc    &  8&\bf{41.33}&     42.62&     44.85&     52.15\\\hline
\gaia        & 32&     25.85&     25.34&\bf{23.78}&     24.40\\\hline
\osm         & 32&     36.73&     32.13&     31.92&\bf{31.19}\\\hline
\uniformb    & 32&     20.35&\bf{19.99}&     20.15&     21.74\\\hline
\uniformc    & 32&     90.01&\bf{73.99}&     78.15&     84.43\\\hline
\gaia        &128&     41.45&\bf{39.42}&     40.32&     40.38\\\hline 
\osm         &128&     35.60&     32.51&\bf{31.57}&     32.24\\\hline 
\uniformb    &128&     41.65&\bf{37.27}&     38.27&     40.23\\\hline 
\uniformc    &128&    184.45&\bf{168.01}&    172.23&    172.67\\\hline  
\end{tabularx}
\end{footnotesize}
\end{table}

\subsubsection{Work Queue Performance: GPU Monolithic Batch Size} We examine performance as a function of the selection of the monolithic batch size. 
The selection of the monolithic batch size, $n_{large}$, has several performance implications.  A large $n_{large}$ will decrease the fraction of queries that \hgpu is able to successfully compute in the first batch round, as $\epsilon$ is selected to find on average $K$ neighbors per $p_i\in D$ (Section~\ref{sec:select_epsilon}). A small value of $n_{large}$ will decrease GPU throughput, where GPU resources may not be sufficiently saturated. We examine the performance impact of the monolithic batch size when we do not reserve any queries for the CPU ($n_{Cwin}=0$). This allows us to observe how performance changes across values of $n_{large}$ in the range $[0.05, 1.0]$. Note that when $n_{large}=1.0$, the GPU can be assigned the entire dataset to process during its first batch round and this will have the effect of starving the CPU of work.

\begin{figure}[!t]
\centering

\subfigure[\uniforma]{
\includegraphics[width=0.22\textwidth, trim={0.4cm 0.4cm 0.4cm 0.4cm}]{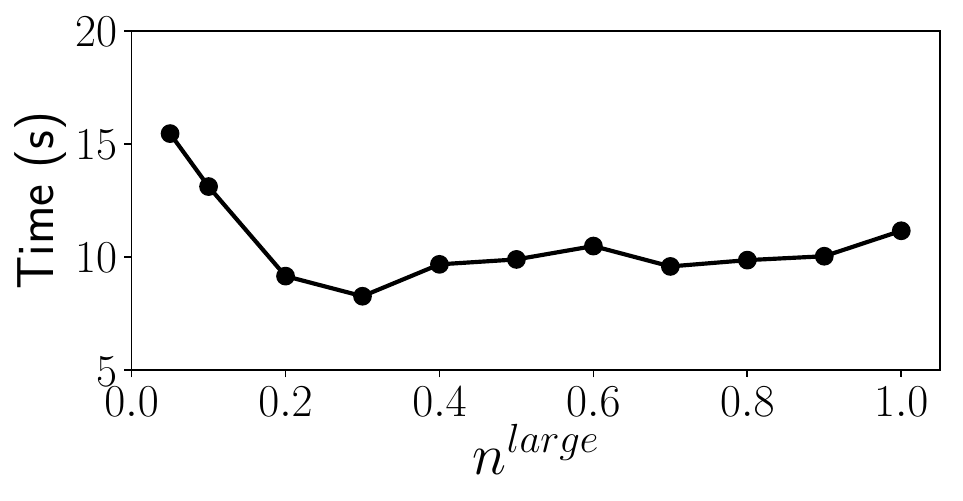}
}
\subfigure[\expoa]{
\includegraphics[width=0.22\textwidth, trim={0.4cm 0.4cm 0.4cm 0.4cm}]{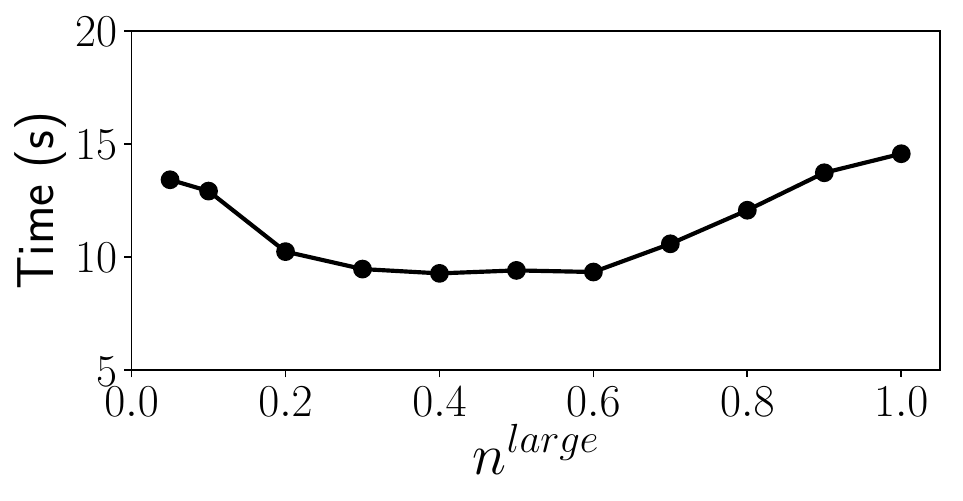}
}

\subfigure[\uniformb]{
\includegraphics[width=0.22\textwidth, trim={0.4cm 0.4cm 0.4cm 0.4cm}]{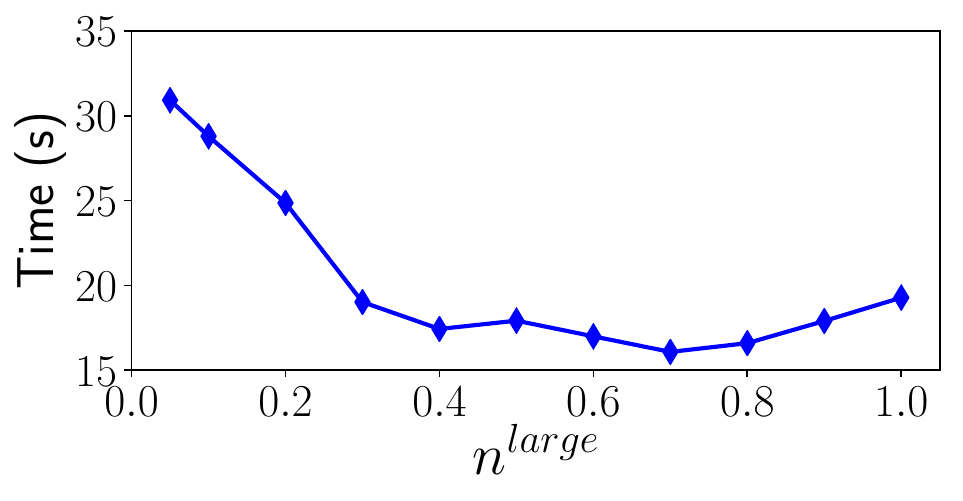}
}
\subfigure[\expob]{
\includegraphics[width=0.22\textwidth, trim={0.4cm 0.4cm 0.4cm 0.4cm}]{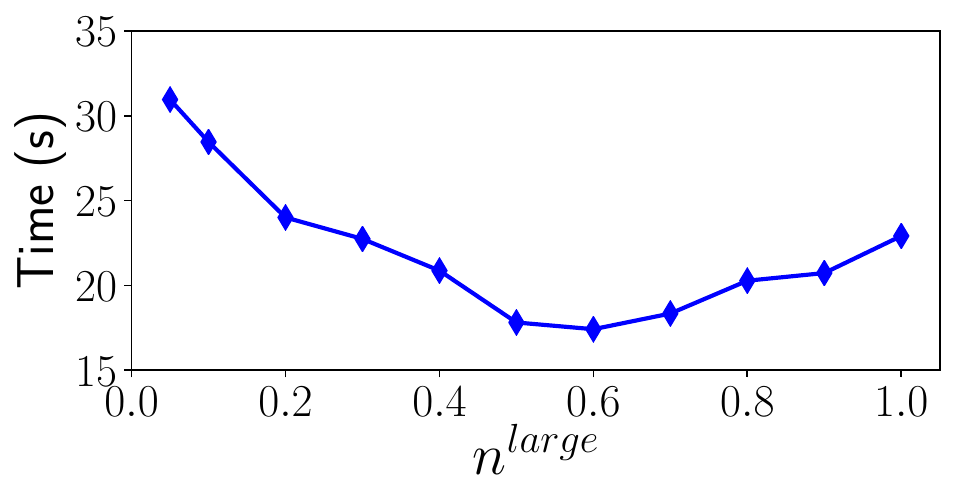}
}

\subfigure[\uniformc]{
\includegraphics[width=0.22\textwidth, trim={0.4cm 0.4cm 0.4cm 0.4cm}]{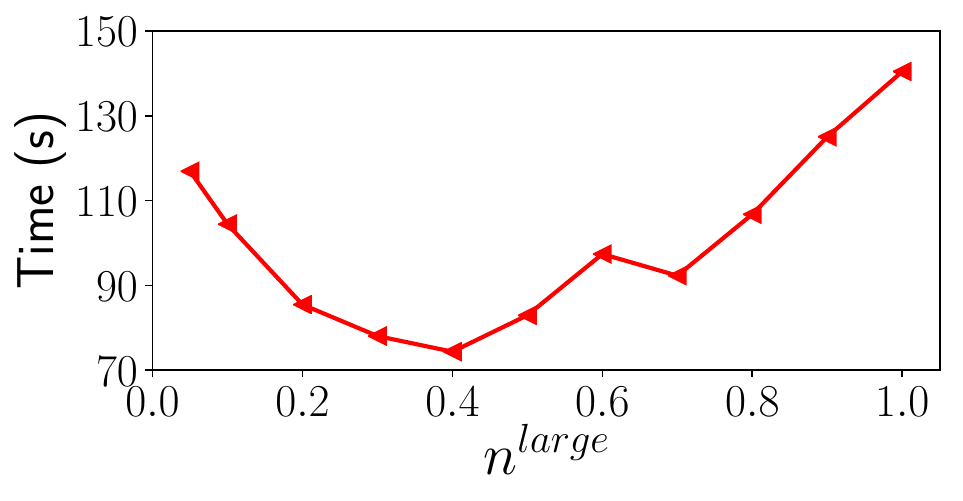}
}
\subfigure[\expoc]{
\includegraphics[width=0.22\textwidth, trim={0.4cm 0.4cm 0.4cm 0.4cm}]{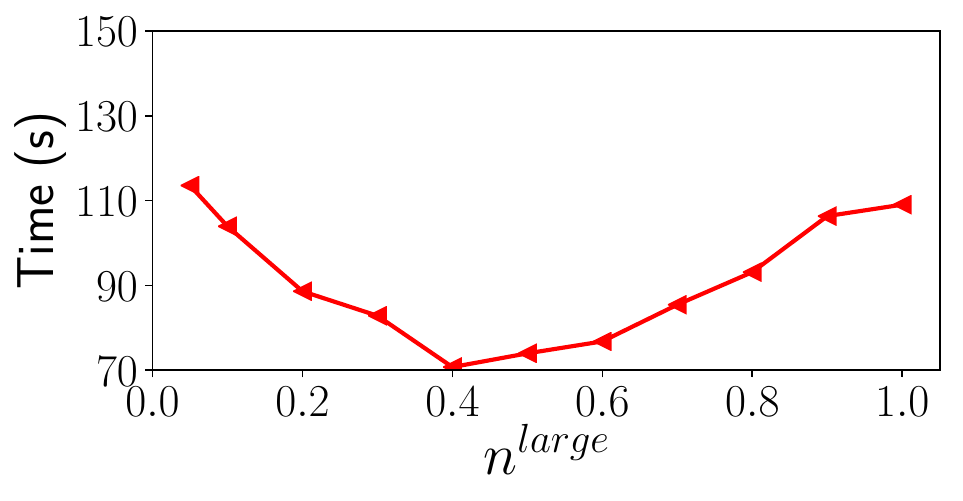}
}

    \caption{\gpu  response time vs. monolithic batch size $n_{large}$ where $K=32$ and $n_{Cwin}=0$. Excepting $n_{large}$ and $n_{Cwin}$, the default parameter values in Table~\ref{tab:params} are used.}
   \label{fig:monolithic_batch}
\end{figure}

Figure~\ref{fig:monolithic_batch}(a),~(c),~and~(e) plot the response time vs. $n_{large}$ on the \uniforma, \uniformb, and \uniformc datasets, respectively, and the exponential datasets are shown in Figure~\ref{fig:monolithic_batch}(b),~(d),~and~(f). For clarity, the uniform and exponential datasets of the same dimensionality are positioned adjacent to each other. From Figure~\ref{fig:monolithic_batch}(a),~(c),~and~(e), we find that $n_{large}$ should not be too small, otherwise GPU resources will not be fully utilized, which is shown by the initial decrease in response time (e.g., comparing $n_{large}=0.05$ and $0.2$ in Figure~\ref{fig:monolithic_batch}(a)). However, on \uniforma, we observe that too large a value of $n_{large}$ will decrease performance. This effect is more pronounced on the \uniformc dataset, where we find that the best value of $n_{large}=0.4$, which shows that \hgpu should be configured with large monolithic batches when computing larger workloads. However, the value should not be too large, otherwise, the CPU will be starved of work, which explains the performance degradation when $n_{large}>0.4$ on \uniformc. Similar results are shown on the exponential datasets in Figure~\ref{fig:monolithic_batch}(b),~(d),~and~(f), so we omit a similar discussion.

Comparing the results in Figure~\ref{fig:monolithic_batch}, $n_{large}$ can be selected in a large range to achieve good performance (the response times are similar between $n_{large}\approx0.3-0.6$). Hence, we select $n_{large}=0.4$ in Table~\ref{tab:params} to achieve a compromise between small (low dimensionality) and large (high dimensionality) workloads.

\subsubsection{Work Queue Performance: Load Balancing}

We determine whether the configuration of the work queue is able to mitigate load imbalance between CPU and GPU components.  Load imbalance is defined as: $|T_{CPU}-T_{GPU}|/T$, where $T_{CPU}$ is the time that the last executing \hcpu rank finishes computation, $T_{GPU}$ is the time that the \hgpu rank finishes computing its last batch, and $T$ is the total response time. 

\begin{figure*}[!t]
\centering
\includegraphics[width=0.65\textwidth, trim={0.4cm 0.4cm 0.4cm 0.4cm}]{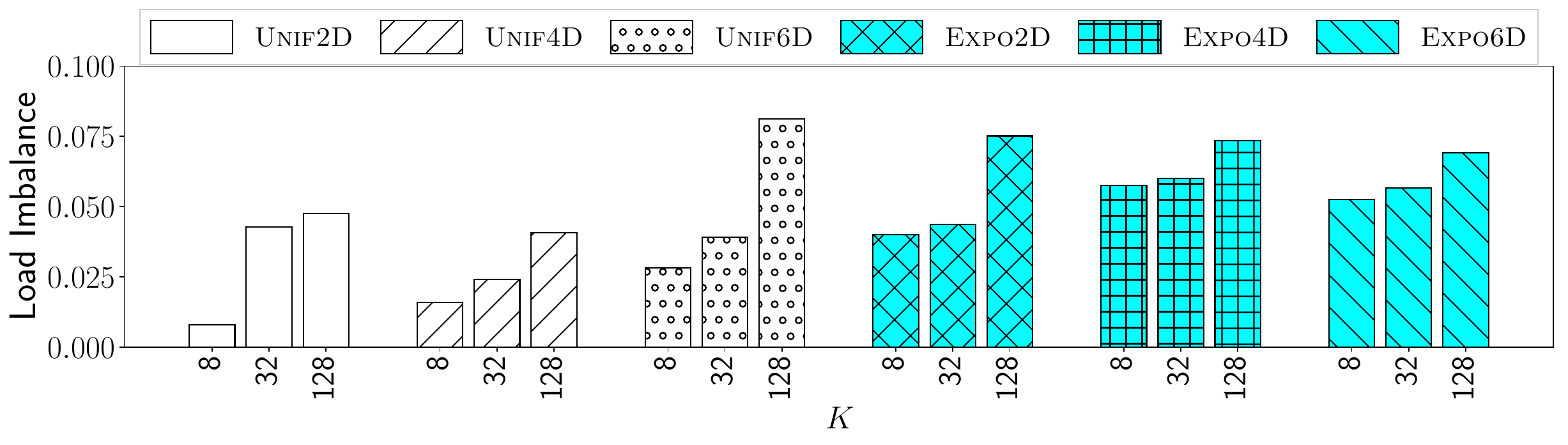}
    \caption{Load imbalance on the uniformly and exponentially distributed datasets for $K=8, 32, 128$. Default parameter values in Table~\ref{tab:params} are used.}
   \label{fig:load_imbalance}
\end{figure*}

Figure~\ref{fig:load_imbalance} shows the load imbalance for $K=8, 32, 128$ on the \classuniform and \classexpo classes of datasets. Across all datasets and values of $K$, we find that the load imbalance is $<10\%$ and generally increases with $K$. As $K$ increases, the total work computed by each batch increases. This enhances the chances of either the CPU or GPU components of \gpu finishing their work at disparate times.  We find that \gpu achieves reasonably good load balancing despite several confounding issues.  To further mitigate load imbalance, the batch sizes could be adjusted based on the value of $K$. For instance, when $K$ is large, smaller batches can be employed. We do not consider this optimization, as it would require an additional parameter that scales as a function of $K$, and it is unlikely to lead to substantial performance gains, as the load imbalance is already within an acceptable range ($<10\%$).

\subsubsection{Quantifying the Number of Failed Queries}\label{sec:failed_queries}
As described in Section~\ref{sec:hgpu}~and~\ref{sec:workqueue}, \hgpu will fail to find some of the query points in each batch. This design decision was made to avoid dynamically increasing the search radius and number of searched cells inside the kernel.\footnote{In an early implementation, we dynamically increased the search radius inside the kernel, but found in preliminary experiments that it led to poor performance due to low warp execution efficiency.} In this section, we examine the number of failed queries on all datasets for selected values of $K$. 

Table~\ref{tab:replay_time} shows the fraction of failed queries computed as the total number of failed queries divided by the total number of attempted queries. The fraction of failed queries ranges from 0--0.342. There are two key observations regarding failed queries described as follows:
\begin{enumerate}
\item Clearly, failed queries are wasted work computed by the GPU. From Table~\ref{tab:replay_time}, up to 34.2\% of queries are wasted.
\item If $\epsilon$ is sufficiently large you can always find the \knn of each query point, and eliminate the wasted work in (1) above. However, this is at the expense of wasted work in another context: many of the query points will find significantly more neighbors than needed, and this increases the overhead of the refinement step, thereby increasing the number of distance calculations.   
\end{enumerate}

\begin{table}[!t]
\centering

\begin{footnotesize}
\caption{The fraction of failed \hgpu queries when executing \gpu, calculated as the total number of failed queries divided by the total number of attempted queries. The default parameter values in Table~\ref{tab:params} are used.}\label{tab:replay_time}
\begin{tabularx}{\columnwidth}{|c|R|R|R|R|R|} \hline
Dataset &$K=8$&$K=32$&$K=96$\\\hline
\gaia        &0.073 &0.164 &0.240\\\hline
\osm         &0.000 &0.000 &0.000  \\\hline
\uniforma    &0.163 &0.135 &0.162\\\hline
\uniformb    &0.217&0.198&0.273\\\hline
\uniformc    &0.270&0.236&0.211\\\hline
\expoa       &0.087&0.058&0.041\\\hline     
\expob       &0.257&0.280&0.296\\\hline
\expoc       &0.342&0.274&0.250\\\hline
\end{tabularx}
\end{footnotesize}
\end{table}

Since the fraction of failed queries is not too large (e.g., a fraction $>0.5$ is likely too large), it indicates that we are reaching a trade-off between not finding too few or too many points. In other words, if the fraction of failed queries was too small, we would expect that we are wasting work refining too many candidate points for the average query point. From Table~\ref{tab:replay_time} we find that on the \osm dataset, we (nearly) always find at least $K$ neighbors for each query point. This indicates on this dataset that a lower value of $\epsilon$ would likely lead to better performance, as the algorithm is refining many more candidate points than needed. We will elaborate on this observation in Section~\ref{sec:epsilon_factor}.

All implementations that use indexes for \knn searches such as the \cpuonly and \kdtree reference implementations will suffer from the refinement overhead described in (2) above. This is because an index cannot guarantee that only $K$ neighbors will be found for a given search. Therefore, this is not a shortcoming of \hgpu, rather it is a consequence of using an index.

\subsubsection{Performance Impact of the Initial Selection of the Search Distance}\label{sec:epsilon_factor}
\begin{figure}[!t]
\centering
\includegraphics[width=0.35\textwidth]{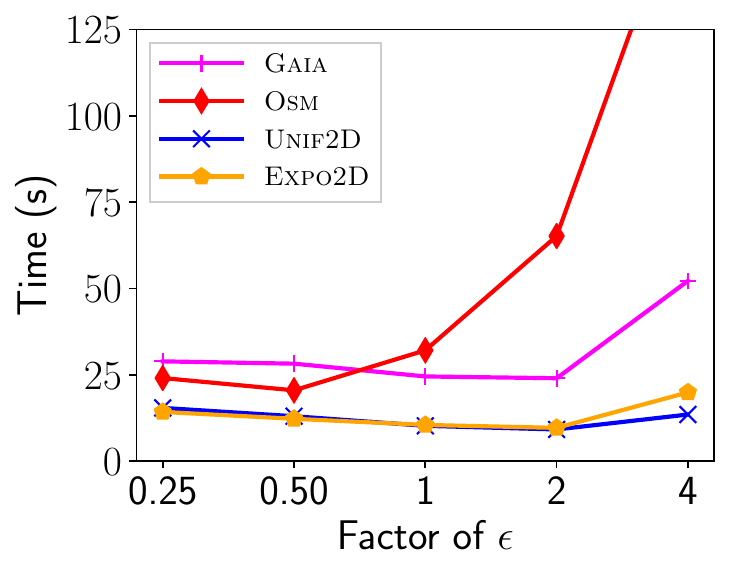}
    \caption{Examining the sensitivity of \gpu to the initial selection of $\epsilon$ on the 2-D datasets, where $K=32$. We execute \gpu with a factor 0.25, 0.5, 2, and 4 of the initial search radius, $\epsilon$, where a value of 1 corresponds to the default $\epsilon$ value.  Default parameter values in Table~\ref{tab:params} are used.}
   \label{fig:epsilon_factor}
\end{figure}

As described in Section~\ref{sec:select_epsilon}, we select an initial search distance, $\epsilon$, that is expected to find at least $K$ neighbors on average for each query point. In this section, we examine how sensitive the performance of \gpu is to variations in this parameter. Figure~\ref{fig:epsilon_factor} plots the response time of \gpu vs. the factor of the initial value of $\epsilon$. In the plot the value of 1 corresponds to the initial/default value of $\epsilon$. We vary the value of $\epsilon$ to be 0.25$\times$--4$\times$ the initial search distance. From Figure~\ref{fig:epsilon_factor}, we find that at a factor 2$\epsilon$, the \gaia, \uniforma, and \expoa datasets achieve slightly better performance than at the default value of $\epsilon$; however, the performance of \osm degrades significantly at $2\epsilon$. Additionally, we find that on \osm, the best performance is achieved at $0.5\epsilon$. Overall, our method for selecting an initial value of $\epsilon$ achieves good performance across all datasets in Figure~\ref{fig:epsilon_factor}.  This supports the semi-analytic geometric justification for the selection of $\epsilon$ outlined in Section~\ref{sec:select_epsilon}.

In Section~\ref{sec:failed_queries} we observed that the \osm dataset has nearly 0 failed queries (Table~\ref{tab:replay_time}). From Figure~\ref{fig:epsilon_factor}, we observe that at a lower value of $\epsilon$, such as $0.5\epsilon$, \osm achieves better performance than at the default $\epsilon$ value. This demonstrates that having nearly no failed queries is an indicator that too many neighbors are being found, which yields additional distance calculations and associated overhead. Also, it demonstrates that having a moderate fraction of query failures is beneficial for performance.

\subsubsection{Temporal Evolution: Re-indexing and Batch Sizes}
At each batch, the \hgpu component of \gpu counts the number of failed queries. If the number of failed queries on the previous batch exceeds 25\% of the number of query points in the batch, the algorithm will re-index with a larger value of $\epsilon$ (Sections~\ref{sec:hgpu}~and~\ref{sec:workqueue}). During this time no queries are computed on the GPU. Table~\ref{tab:reindex_overhead} shows the fraction of the total response time spent re-indexing across all datasets for various values of $K$.\footnote{A non-integral average number of times \gpu needs to re-index is counterintuitive. Therefore, for clarity, we only used a single time trial in this experiment.} Values in parentheses indicate the number of times \hgpu re-indexed. From Table~\ref{tab:reindex_overhead} we find that \hgpu will sometimes not re-index at all (\osm for all values of $K$), and will re-index up to 3 times (\expob, $K=8$). The median number of times the algorithm will re-index is one. Furthermore, re-indexing accounts for 0\%--8.3\% of the total response time. In the majority of cases, less than 3\% of the total response time is spent re-indexing. This demonstrates that re-indexing overhead does not significantly degrade the GPU's query throughput.

\begin{table}[!t]
\centering
\begin{footnotesize}
\caption{The fraction of the total response time spent re-indexing \hgpu when $\epsilon$ is expanded for various values of $K$. Values in parentheses show the number of times \hgpu needed to re-index. The default parameter values in Table~\ref{tab:params} are used.}\label{tab:reindex_overhead}
\begin{tabularx}{\columnwidth}{|c|R|R|R|R|R|} \hline
Dataset &$K=8$&$K=32$&$K=96$\\\hline
\gaia        &0.039 (1)&0.026 (1)&0.029 (1)\\\hline
\osm         &0.000 (0)&0.000 (0)&0.000 (0)\\\hline
\uniforma    &0.030 (1)&0.024 (1)&0.021 (1)\\\hline
\uniformb    &0.028 (1)&0.018 (1)&0.017 (1)\\\hline
\uniformc    &0.015 (1)&0.009 (1)&0.005 (1)\\\hline
\expoa       &0.061 (1)&0.027 (1)&0.020 (1)\\\hline     
\expob       &0.083 (3)&0.034 (2)&0.021 (2)\\\hline
\expoc       &0.036 (2)&0.017 (2)&0.011 (2)\\\hline
\end{tabularx}
\end{footnotesize}
\end{table}

\begin{figure}[!t]
\centering
\subfigure[\uniformb, $K=8$]{
\includegraphics[width=0.4\textwidth]{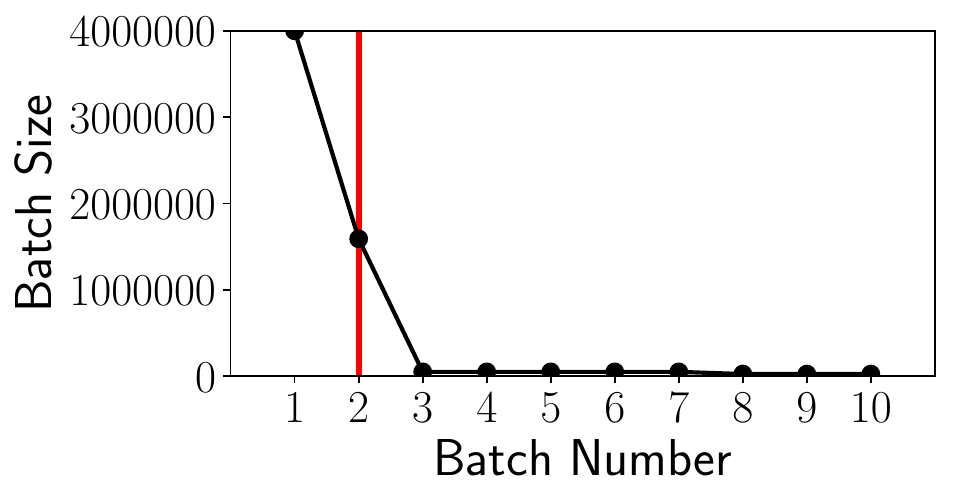}
}

\subfigure[\expob, $K=8$]{
\includegraphics[width=0.4\textwidth]{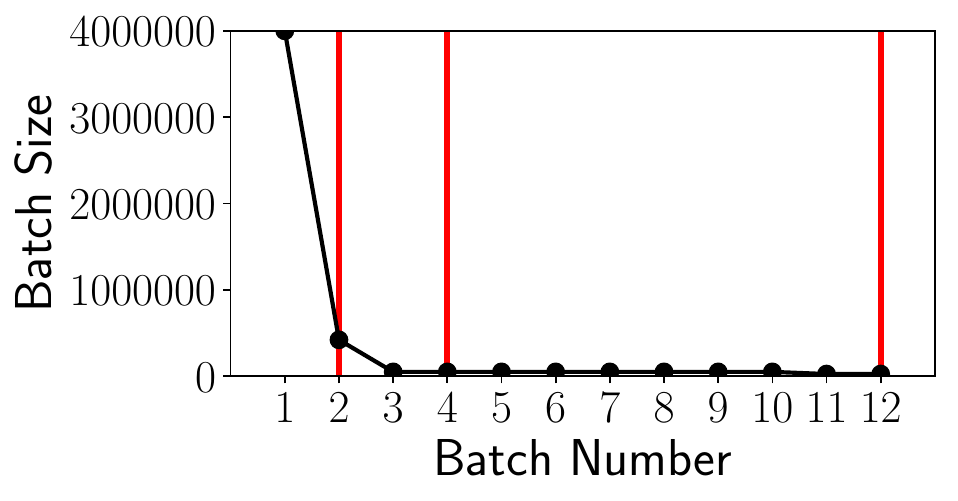}
}
    \caption{\hgpu batch size vs. batch number for (a)~\uniformb and (b)~\expob, where $K=8$. The vertical red lines correspond to the batch number that triggered re-indexing the dataset.}
   \label{fig:temporal_evolution_batches}
\end{figure}

Using the same experiments shown in Table~\ref{tab:reindex_overhead}, Figure~\ref{fig:temporal_evolution_batches} plots the total number of queries assigned to \hgpu at each batch on \uniformb and \expob using $K=8$. The vertical red lines correspond to the batch number that triggered recomputing the index. The figure shows the temporal evolution of \gpu indicating that large workloads are initially assigned to the GPU at batches 1 and 2, which decrease to ensure that low load imbalance occurs between CPU and GPU components of the algorithm. 

\subsubsection{Characterization of the \hgpu Search Distance Expansion}\label{sec:eps_expansion_problem}
Recall from Section~\ref{sec:workqueue} that \hgpu expands $\epsilon$ while computing batches of query points.  We examine the performance of the \hgpu expansion of $\epsilon$ through an experiment using \gpuonly.  As discussed in Section~\ref{sec:implementations}, the \gpuonly implementation is used for comparison purposes, but is not designed for standalone \knn searches. In this section, we also elaborate on the reasoning for the unsuitability of \gpuonly (that uses \hgpu) as a standalone \knn search algorithm.

\begin{figure}[!t]
\centering
\subfigure[]{
\includegraphics[width=0.32\textwidth]{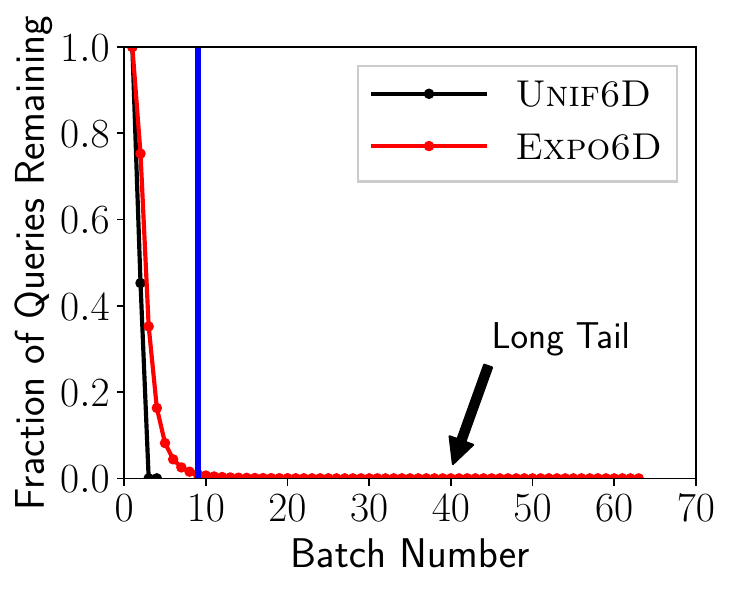}
}

\subfigure[]{
\includegraphics[width=0.32\textwidth]{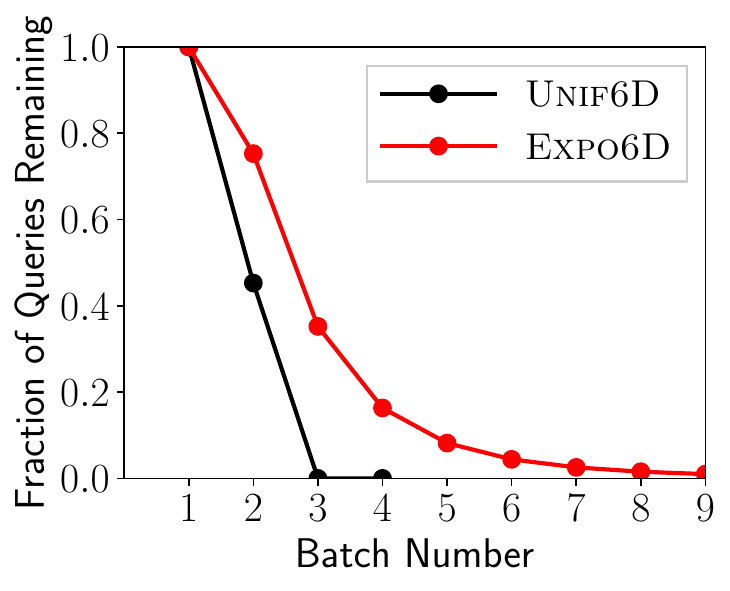}
}
    \caption{The fraction of queries in $D$ that remain to be computed vs. the batch number when executing \gpuonly on the \uniformc and \expoc datasets. (a) The \uniformc dataset is computed in 4 batches, whereas the \expoc dataset requires 64 batches to find the \knn of all points. The vertical line at 9 batches denotes where 99\% of the points in $D$ have been computed on the \expoc dataset. (b) The first 9 batches in (a).}
   \label{fig:gpuonly_batches}
\end{figure}

We execute \gpuonly on the \uniformc and \expoc datasets. Figure~\ref{fig:gpuonly_batches}(a) plots the fraction of queries remaining to be processed vs. the batch number. With each successive batch, the number of queries left to be computed decreases, but the number of successfully solved queries is dependent on the search distance, $\epsilon$, and the data distribution. For example, when the batch number is 1 (the first batch), then 100\% of the queries still need to be computed. From Figure~\ref{fig:gpuonly_batches}(a) we observe that on the \uniformc dataset only 4 batches are required to find the \knn of all query points in the dataset. However, on the \expoc dataset, 64 batches are needed to compute the \knn of all points in the dataset. However, on the \expoc dataset, the \knn of 99\% of the points in the dataset are found after only computing 9 batches (denoted by the vertical blue line). The remaining 1\% of the data points require computing 64-9=55 additional batches. Figure~\ref{fig:gpuonly_batches}(b) shows Figure~\ref{fig:gpuonly_batches}(a) but constrained to the first 9 batches. From Figure~\ref{fig:gpuonly_batches}(b), we find that if we exclude the remaining 1\% of data points that require executing an additional 55 batches (the long tail in Figure~\ref{fig:gpuonly_batches}(a)), then the fraction of queries remaining to be computed by \expoc is reasonable. For comparison, when 4 batches have been computed, the \knn have been found for all points in the \uniformc dataset, and only 16.3\% of the points are left to compute on the \expoc dataset.

In general, we find that when the data are uniformly distributed, the selection of the initial value of $\epsilon$ is able to find the \knn of a large fraction of the data points with few expansions of $\epsilon$. In contrast, when the data are exponentially distributed, the mean value of $\epsilon$ needed to find $K$ neighbors on average ($\epsilon_{min}$) is largely unsuitable for those data points in sparse regions. Thus, the \gpuonly algorithm  needs to continually expand $\epsilon$ to find the points in these sparse regions, and this may require a prohibitive number of $\epsilon$ expansions that degrades performance.

To remedy this problem in a GPU-only implementation, a non-linear expansion of $\epsilon$ could be employed to avoid the long tail exhibited in Figure~\ref{fig:gpuonly_batches}(a) (recall that $\epsilon$ is expanded by $\epsilon_{min}/2$ when \hgpu fails to find the \knn of at least 25\% of the query points in the previous batch). As an alternative, a simple brute force search could be employed when a small fraction of points in the dataset have not found their \knn (e.g., similarly to \expoc in Figure~\ref{fig:gpuonly_batches}, where 99\% of the data points find their respective \knn in 9 batches, then only 1\% of the data are left to find their respective \knn). In this case, a brute force search will not be prohibitive to performance. However, we reiterate that \hgpu expects those points in the low density regions to be found by \hcpu; therefore, we do not use these solutions, as their implementation would not be executed when using \gpu.

\begin{figure*}[!t]
\centering
\subfigure[\uniforma]{
\includegraphics[width=0.3\textwidth]{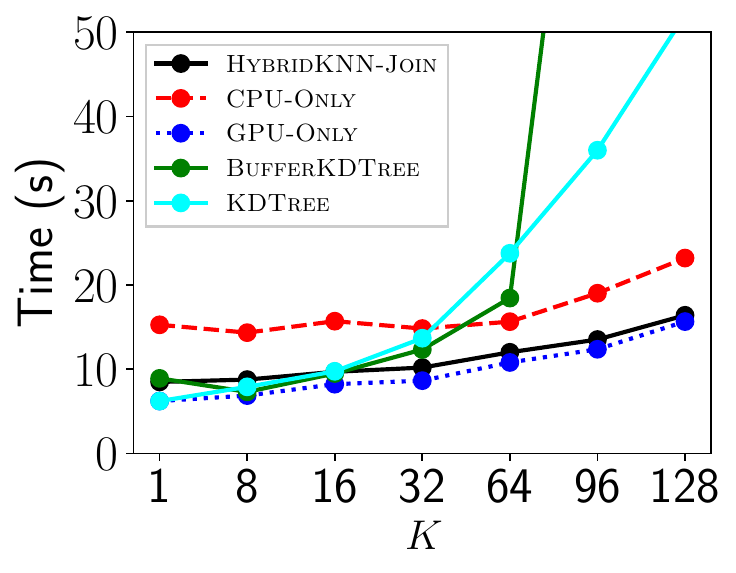}
}
\subfigure[\uniformb]{
\includegraphics[width=0.3\textwidth]{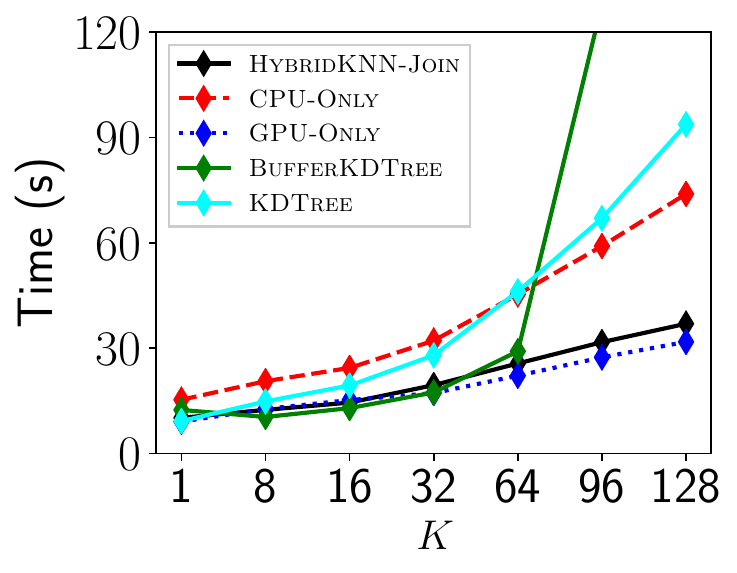}
}
\subfigure[\uniformc]{
\includegraphics[width=0.3\textwidth]{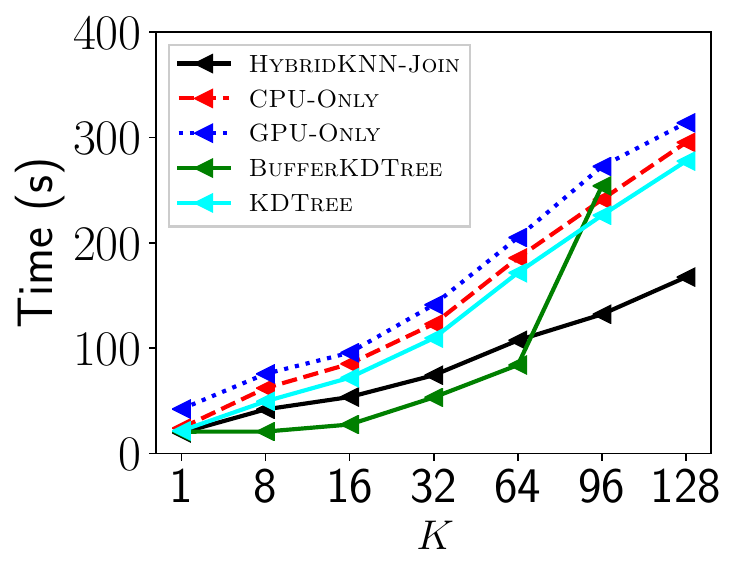}
}

\subfigure[\expoa]{
\includegraphics[width=0.3\textwidth]{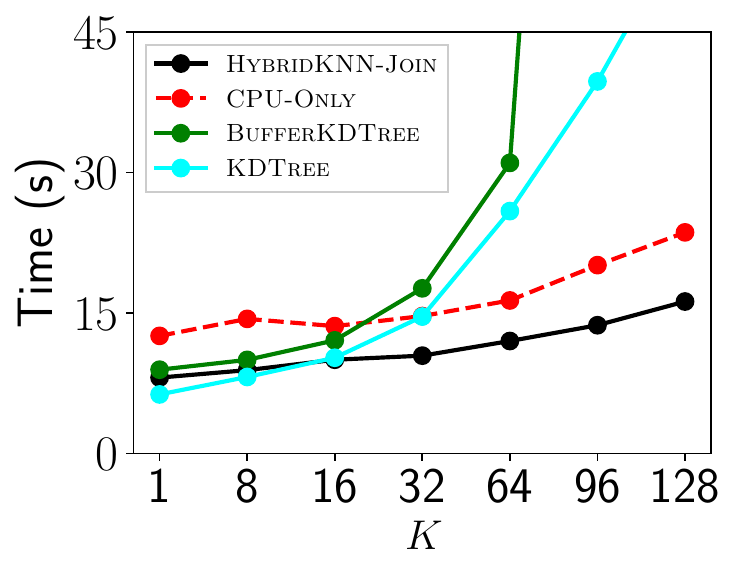}
}
\subfigure[\expob]{
\includegraphics[width=0.3\textwidth]{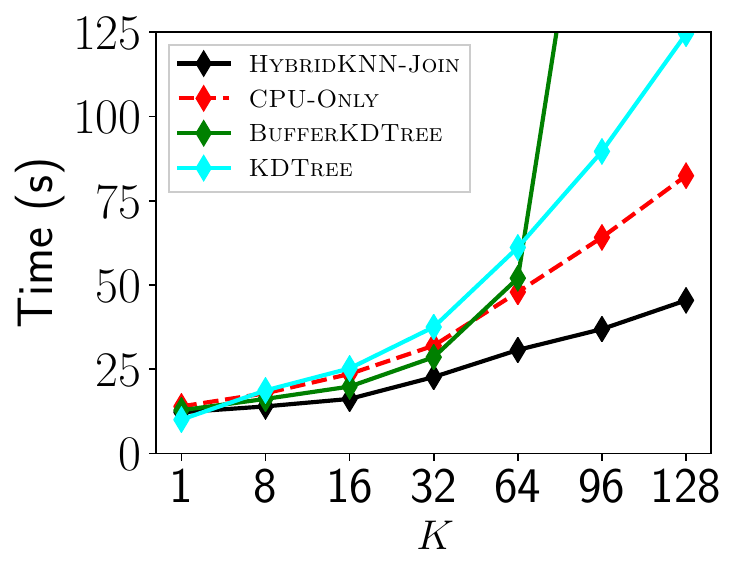}
}
\subfigure[\expoc]{
\includegraphics[width=0.3\textwidth]{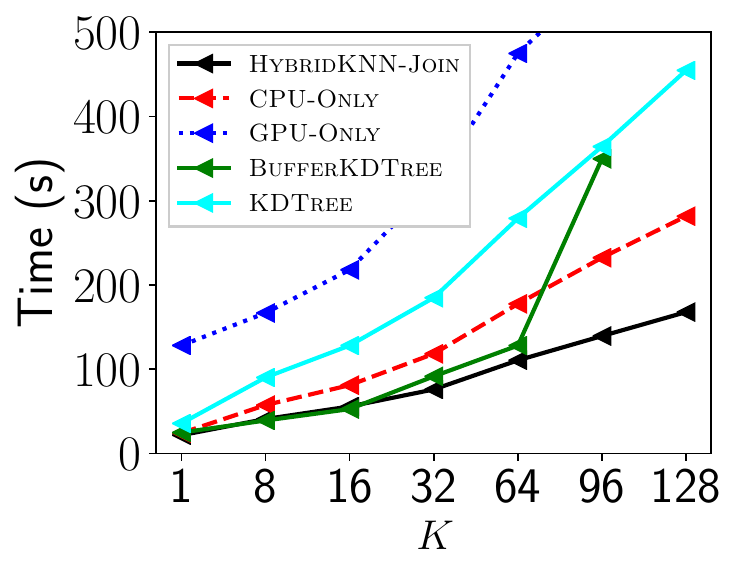}
}
    \caption{Response time vs. $K$ comparing \gpu, \cpuonly, \gpuonly, \bufferkdtree, and \kdtree on (a)--(c) uniformly, and (d)--(f) exponentially distributed datasets. \gpu is configured with the default parameter values in Table~\ref{tab:params}. In (d)~and~(e), we do not plot \gpuonly, as the total response times are much larger than the other implementations due to the explanation given in Section~\ref{sec:eps_expansion_problem}. }
   \label{fig:time_vs_K_synthetic}
\end{figure*}

Figure~\ref{fig:time_vs_K_synthetic}(a)--(c)  plots the response time vs. $K$ on the \classuniform datasets, and the \classexpo datasets are shown in Figure~\ref{fig:time_vs_K_synthetic}(d)--(f). 
Despite the performance drawback of the \gpuonly implementation on the \classexpo datasets described above, we find that on the uniformly distributed datasets, the \gpuonly implementation is very efficient. In fact, we find that \gpuonly slightly outperforms \gpu across all values of $K$ on the \uniforma dataset. Because the dataset contains 2-D points (the smallest workload), the slight overheads and load imbalance between architectures are observable on this dataset. In contrast, on the \uniformc dataset (a larger workload), we find that \gpu outperforms \gpuonly.

\subsubsection{Comparison of \cpuonly and \kdtree}
We compare the performance of the standalone multi-core CPU implementations, \cpuonly and \kdtree. Recall that \cpuonly is parallelized using MPI processes and \kdtree is parallelized using threads. From Figure~\ref{fig:time_vs_K_synthetic}, we observe that the on the synthetic datasets, the performance of \cpuonly degrades gracefully with increasing $K$. In contrast, the performance of \kdtree degrades much more significantly with $K$ on some datasets, such as \uniforma and \expoa. 

\begin{figure}[!t]
\centering
\subfigure[\gaia (2-D)]{
\includegraphics[width=0.3\textwidth]{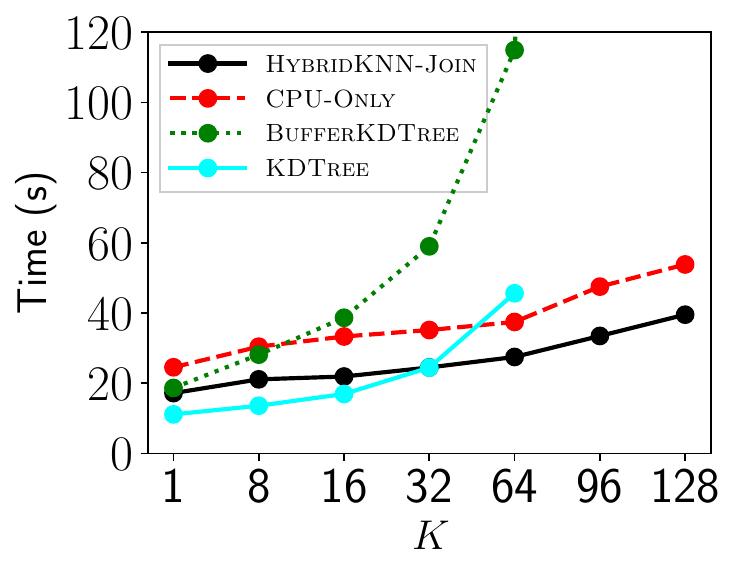}
}

\subfigure[\osm (2-D)]{
\includegraphics[width=0.3\textwidth]{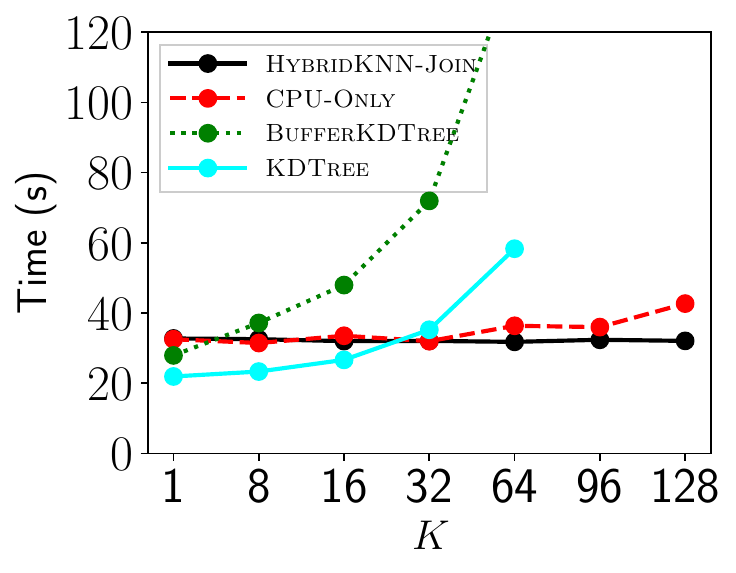}
}
    \caption{Response time vs. $K$ on 2-D real-world datasets. Default parameter values in Table~\ref{tab:params} are used.}
   \label{fig:time_vs_K_real_world}
\end{figure}

Figure~\ref{fig:time_vs_K_real_world} plots the response time vs. $K$ on the real-world datasets. These datasets are 2-D with $|D|=2.5\times10^7$ points.  We note on this experiment that \kdtree failed to execute on $K\geq96$ on both \gaia and \osm. We find that \kdtree outperforms \cpuonly when $K\lesssim 32$, whereas \cpuonly outperforms \kdtree when $K>32$.  Based on the performance of \kdtree in Figure~\ref{fig:time_vs_K_synthetic}, we expect the response time of \kdtree to increase significantly at $K\geq 96$.

\begin{figure}[!t]
\centering
\includegraphics[width=0.35\textwidth, trim={0.4cm 0.4cm 0.4cm 0.4cm}]{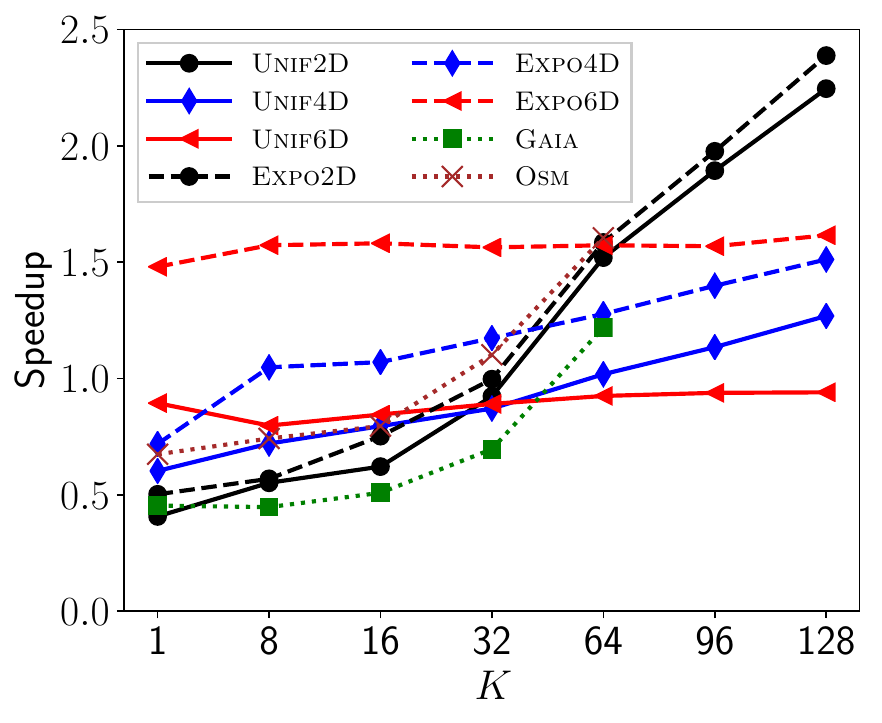}
    \caption{Speedup of \cpuonly over \kdtree vs. $K$ on all datasets shown in Figures~\ref{fig:time_vs_K_synthetic}~and~\ref{fig:time_vs_K_real_world}.}
   \label{fig:speedup_cpuonly_kdtree}
\end{figure}

To more clearly observe the performance differences between \cpuonly and \kdtree, Figure~\ref{fig:speedup_cpuonly_kdtree}  plots the speedup of \cpuonly over \kdtree on all datasets. We find that \cpuonly is much more efficient on the larger values of $K$ on the \uniforma, \expoa, and \expob datasets. Additionally, since these two multi-core CPU reference implementations have varying performance characteristics, they are able to provide a more comprehensive comparison to \gpu (see Section~\ref{sec:hybrid_vs_cpu}).

In general, we find that \cpuonly is mostly competitive with or outperforms \kdtree (Figure~\ref{fig:speedup_cpuonly_kdtree}). This indicates that the use of MPI to parallelize \hcpu and \cpuonly is not prohibitive to the performance of the algorithms. If MPI were prohibitive, we would expect that \kdtree would achieve better performance relative to \cpuonly.

\subsubsection{Comparison of \gpu to \cpuonly and \kdtree}\label{sec:hybrid_vs_cpu}

We compare the performance of \gpu to the two multi-core CPU implementations, \cpuonly and \kdtree. From Figure~\ref{fig:time_vs_K_synthetic}, across all datasets and values of $K$, \gpu outperforms \cpuonly. This demonstrates that the use of the CPU and GPU in the hybrid algorithm does not degrade performance for any value of $K$. Comparing \gpu to \kdtree, we find that there are a few cases where \kdtree outperforms \gpu on lower values of $K$. For example, on \uniforma and \expoa, \kdtree slightly outperforms \gpu when $K\leq8$. Overall, \gpu yields a reasonable performance gains over \kdtree.

On the \gaia dataset (Figure~\ref{fig:time_vs_K_real_world}(a)), \gpu outperforms \cpuonly across all values of $K$. On the \osm dataset (Figure~\ref{fig:time_vs_K_real_world}(b)), \gpu and \cpuonly have nearly equal performance from $1<K\leq32$; however, when $K>32$, \gpu outperforms \cpuonly. We find that \kdtree outperforms \gpu on \gaia and \osm when $K\leq16$. 

It is interesting to note that on \gaia and \osm, \kdtree outperforms \cpuonly on low values of $K$, whereas \cpuonly performs better on the higher values of $K$ (we observed similar behavior on some of the synthetic datasets in Figure~\ref{fig:time_vs_K_synthetic}). This suggests that \gpu could be equipped with different algorithms, such as using \kdtree instead of the ANN-based \cpuonly implementation for lower values of $K$. While this is beyond the scope of this work, algorithm selection as a function of input parameters is an interesting research direction for hybrid CPU/GPU algorithms.

\begin{figure}[!t]
\centering
\subfigure[]{
\includegraphics[width=0.35\textwidth, trim={0.4cm 0.4cm 0.4cm 0.4cm}]{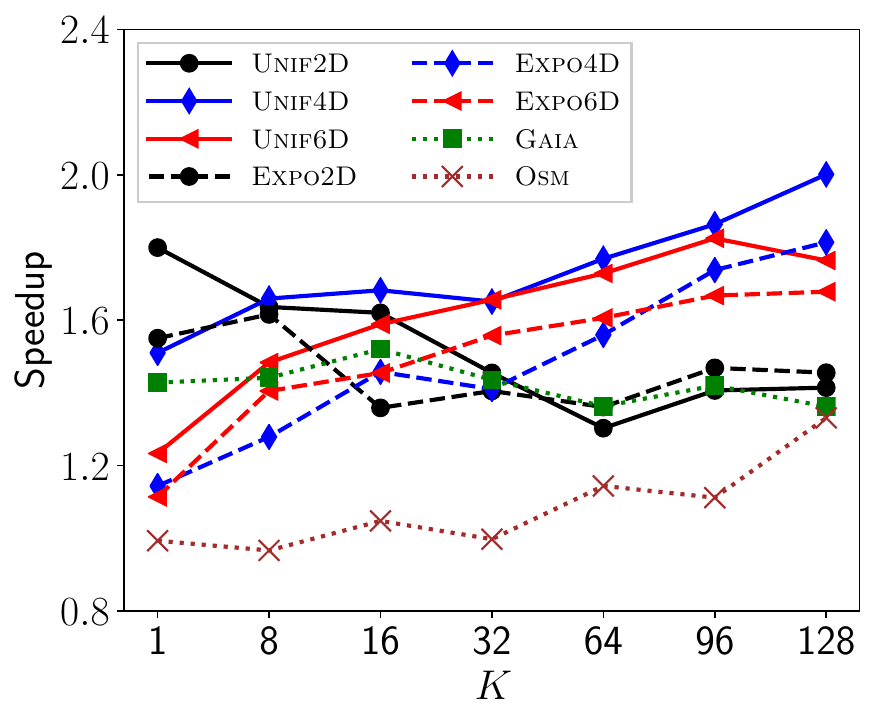}
}

\subfigure[]{
\includegraphics[width=0.35\textwidth, trim={0.4cm 0.4cm 0.4cm 0.4cm}]{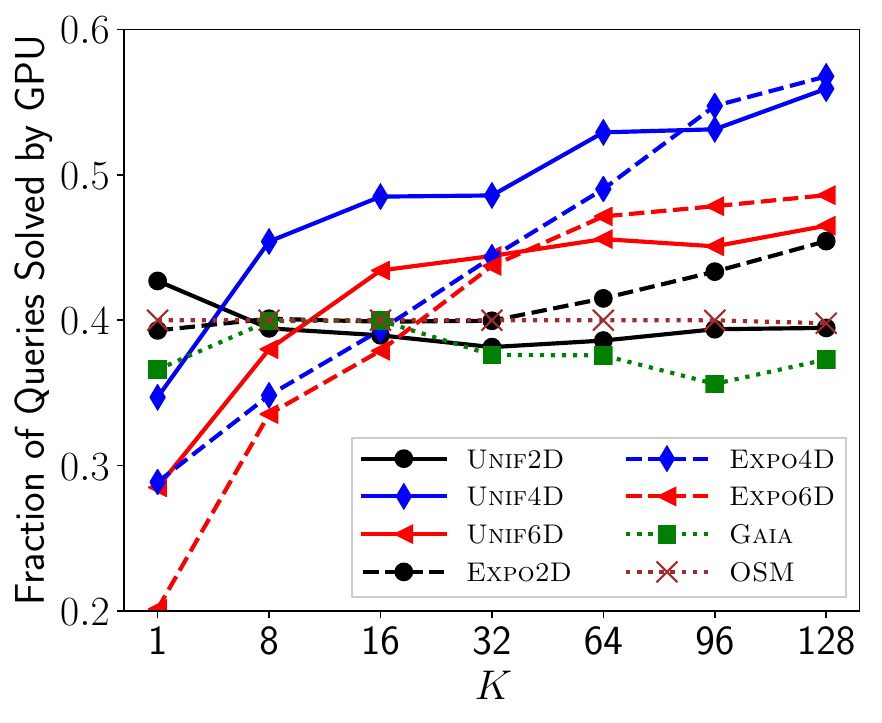}
}
	
    \caption{(a) Speedup of \gpu over \cpuonly vs. $K$ on all datasets in Figures~\ref{fig:time_vs_K_synthetic}~and~\ref{fig:time_vs_K_real_world}. With the exception of the 2-D datasets, the speedup roughly increases with dimensionality and $K$. (b) Fraction of query points solved by \hgpu. Excepting 2-D datasets, the GPU computes a larger fraction of $D$ with increasing $K$.}
   \label{fig:speedup_vs_K_synthetic}
\end{figure}

Figure~\ref{fig:speedup_vs_K_synthetic}(a) plots the speedup of \gpu over \cpuonly for all datasets in Figures~\ref{fig:time_vs_K_synthetic}~and~\ref{fig:time_vs_K_real_world}. The plot demonstrates that the performance advantage of \gpu is greater when $K$ is large or when the dimensionality increases (excepting the 2-D datasets). Figure~\ref{fig:speedup_vs_K_synthetic}(b) shows that as $K$ increases, the fraction of queries solved by \hgpu also increases on most datasets. Note that while the fraction of $D$ computed by \hgpu is generally $<$50\%, \hgpu is computing the queries with the greatest amount of work in the denser data regions of the \gaia, \osm, and \classexpo datasets (by definition, the \classuniform  datasets  have constant density across the data space).

\subsubsection{Comparison of \gpu to \bufferkdtree}
\begin{figure}[!t]
\centering
\includegraphics[width=0.35\textwidth, trim={0.4cm 0.4cm 0.4cm 0.4cm}]{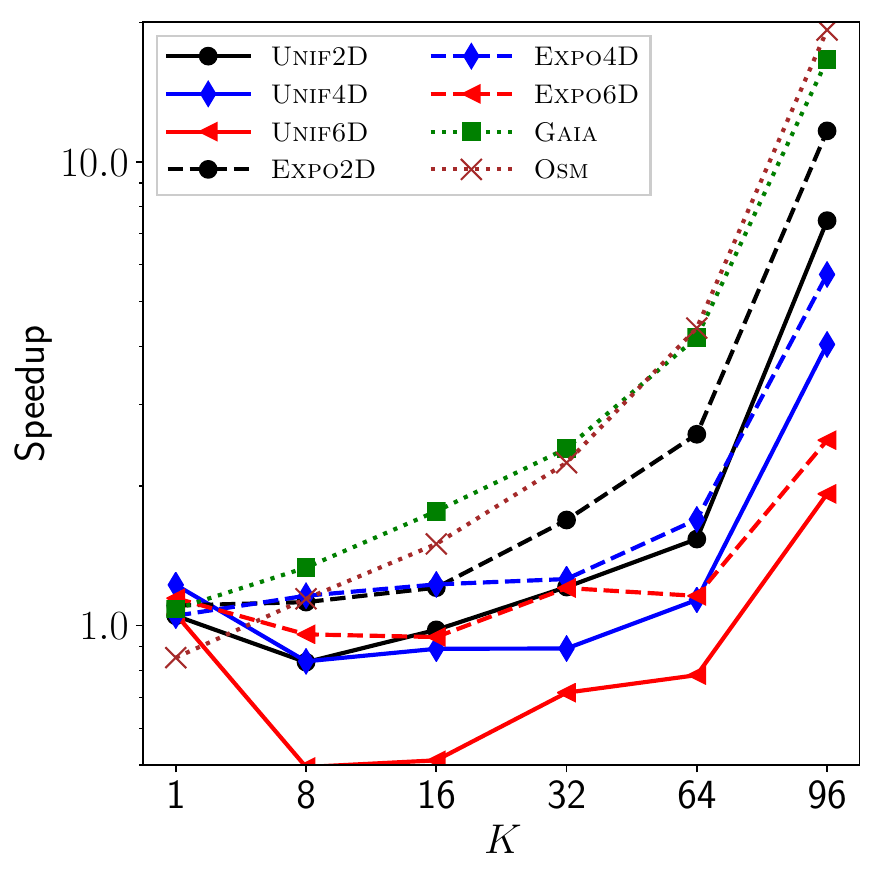}
    \caption{Speedup of \gpu over \bufferkdtree vs. $K$ on all datasets shown in Figures~\ref{fig:time_vs_K_synthetic}~and~\ref{fig:time_vs_K_real_world}.}
   \label{fig:speedup_bufferkdtree_vs_K}
\end{figure}

Figure~\ref{fig:speedup_bufferkdtree_vs_K} plots the speedup of \gpu over the GPU algorithm \bufferkdtree vs. $K$ on all datasets shown in Figures~\ref{fig:time_vs_K_synthetic}~and~\ref{fig:time_vs_K_real_world}, corresponding to the synthetic and real-world datasets, respectively. 

We find that on most synthetic datasets, \gpu achieves a speedup over \bufferkdtree. However, \bufferkdtree outperforms \gpu on some smaller values of $K$ (e.g., on the \uniforma and \uniformb datasets), and outperforms \gpu on \uniformc when $K\leq 64$. Overall, we find that \gpu significantly outperforms \bufferkdtree on the larger values of $K$. 

On the real-world datasets, \gaia and \osm, the performance gains over \bufferkdtree are more pronounced than on the synthetic datasets. On the \gaia dataset, the speedup ranges from 1.08--16.59$\times$, and on the \osm dataset the speedup ranges from 0.85--19.24$\times$. Therefore, there is only one case at $K=1$ on \osm where \gpu achieves a slowdown relative to \bufferkdtree. 

\section{Discussion \& Conclusions}\label{sec:conclusions}
Many of the GPU \knn works address high-dimensionality \cite{Garcia2010,arefin2012gpu,jian2013parallel,komarov2014fast}. Here, we advance a hybrid approach for low-dimensionality that exploits the relative strengths of the CPU and GPU architectures. GPU \knn algorithms are less likely to achieve significant performance gains in low dimensionality due to highly efficient CPU algorithms, such as ANN~\cite{arya1998optimal}.  

We consider the throughput-oriented GPU vs. the low-latency CPU. Our strategy assigns large batches to the GPU to maintain high throughput, while the CPU ranks are assigned smaller chunks of work. We largely mitigate load imbalance and starvation by reducing the batch size assigned to the GPU depending on the number of completed queries, and reserving queries for the CPU. The work queue allows new advances in GPU- and CPU-only algorithms to be substituted into the framework to further improve performance.  More broadly, the work queue could be used as a general technique to address other CPU/GPU algorithms with data-dependent performance characteristics.

\gpu yields reasonable performance gains over the reference implementations. We find that the speedup over the parallel CPU approach (\cpuonly) is $<2\times$; however, from Figure~\ref{fig:speedup_vs_K_synthetic}(a), we clearly observe that the speedup of \gpu is expected to be greater at higher values of $K$ (and potentially dimensionality) than the scenarios examined in this paper. Similarly, we find that \gpu outperforms the GPU reference implementation, \bufferkdtree, on most scenarios (Figure~\ref{fig:speedup_bufferkdtree_vs_K}).  

An overall observation from this exercise is that hybrid algorithms are difficult to design. Since the performance of each CPU and GPU algorithm largely varies due to input parameters ($K$) and data properties, it is challenging to design an algorithm that will outperform or achieve comparable performance to all other CPU-only or GPU-only reference implementations.

We found that there are some experimental scenarios where \gpu yields a slowdown compared to some of the reference implementations (e.g., the \kdtree implementation on low values of $K$). Since all \knn algorithms have particular performance niches, hybrid algorithms could be developed to include algorithm selection as a function of several parameters, such as $K$, data dimensionality, and data distribution. This would allow hybrid algorithms to achieve better performance over a wider range of scenarios. We leave this research direction for future work.

A recent trend in computer architecture is the use of GPUs in clusters. For example, each compute node in the Summit supercomputer at Oak Ridge National Laboratory contains six Nvidia Volta GPUs~\cite{Summit}. The work queue proposed in this paper could be used to distribute work to multiple GPUs within a single node. Furthermore, the work queue is a good design for workloads that vary based on data distribution, and could be applied to other spatial search algorithms used for data analysis, such as similarity searches~\cite{kalashnikov2013}, and DBSCAN clustering~\cite{ester1996density,Bohm:2000:HPC:354756.354832}.  Based on our experiments, the work queue is able to achieve good load balancing between the CPU and GPU, and therefore, we would expect to achieve good load balancing between multiple GPUs. The work queue would only need to be reconfigured to incorporate several GPU consumers. Interestingly, on fat-nodes like those in Summit, the use of the CPU in hybrid algorithms would become less important, since the computational throughput of several GPUs would be much higher than the CPUs in the system. Another interesting design is partitioning the input dataset across the global memory of multiple GPUs to enable larger datasets to be processed. New interconnects such as NVLink~\cite{2017Foley} enable direct GPU-to-GPU communication, thus obviating slower main-memory accesses orchestrated by the host.

\section*{Acknowledgments}
This material is based upon work supported by the National Science Foundation under Grant No. 1849559.

\bibliographystyle{elsarticle-num}

 \begin{SCfigure}[50][th]
    \includegraphics[width=1in,height=1.25in,clip,keepaspectratio]{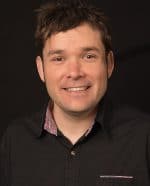}
    \captionsetup{labelformat=empty, width=1.05\textwidth}
    \caption{\footnotesize{ \textbf{Michael Gowanlock} is an assistant professor in the School of Informatics, Computing, \& Cyber Systems at Northern Arizona University. He received a Ph.D. in computer science at the University of Hawai`i at M\=anoa. He was a postdoctoral associate at MIT Haystack Observatory. His research interests include parallel data-intensive computing, and astronomy.}}
  \end{SCfigure}

\end{document}